\documentclass[sigconf]{acmart}
\AtBeginDocument{%
  }

\setcopyright{acmlicensed}
\copyrightyear{2026}
\acmYear{2026}
\setcopyright{cc}
\setcctype{by}
\acmConference[KDD '26]{Proceedings of the 32nd ACM SIGKDD Conference on Knowledge Discovery and Data Mining V.2}{August 09--13, 2026}{Jeju Island, Republic of Korea}
\acmBooktitle{Proceedings of the 32nd ACM SIGKDD Conference on Knowledge Discovery and Data Mining V.2 (KDD '26), August 09--13, 2026, Jeju Island, Republic of Korea}
\acmDOI{10.1145/3770855.3818910}
\acmISBN{979-8-4007-2259-2/2026/08}

\usepackage{multirow}
\usepackage{bm}
\usepackage[table]{xcolor}
\usepackage{cuted}
\usepackage{subcaption}
\usepackage{algorithm}
\usepackage{algpseudocode}
\begin{document}

\title[ECGFlowCMR]{ECGFlowCMR: Pretraining with ECG-Generated Cine CMR Helps
Cardiac Disease Classification and Phenotype Prediction}



\author{Xiaocheng Fang}
\affiliation{%
  \institution{School of Intelligence Science and Technology, Peking University}
  \city{Beijing}
  \country{China}
}
\additionalaffiliation{%
  \institution{State Key Laboratory of General Artificial Intelligence, Peking University}
  \city{Beijing}
  \country{China}
}
\email{fangxiaocheng162@gmail.com}

\author{Zhengyao Ding}
\affiliation{%
  \institution{Polytechnic Institute of Zhejiang University, Zhejiang University}
  \city{Hangzhou}
  \country{China}
}
\email{zhengyao.ding@zju.edu.cn}

\author{Guangkun Nie}
\authornotemark[1]
\affiliation{%
  \institution{School of Intelligence Science and Technology, Peking University}
  \city{Beijing}
  \country{China}
}
\email{nieguangkun@stu.pku.edu.cn}

\author{Jieyi Cai}
\affiliation{%
  \institution{Institute of Computing Technology, University of the Chinese Academy of Sciences}
  \city{Beijing}
  \country{China}
}
\email{caijieyi055@gmail.com}

\author{Yujie Xiao}
\affiliation{%
  \institution{National Institute of Health Data Science, Peking University}
  \city{Beijing}
  \country{China}
}
\email{xiaoyujie@stu.pku.edu.cn}

\author{Bo Liu}
\authornotemark[1]
\affiliation{%
  \institution{School of Intelligence Science and Technology, Peking University}
  \city{Beijing}
  \country{China}
}
\email{liubo2022@stu.pku.edu.cn}

\author{Jiarui Jin}
\authornotemark[1]
\affiliation{%
  \institution{School of Intelligence Science and Technology, Peking University}
  \city{Beijing}
  \country{China}
}
\email{jrjin25@stu.pku.edu.cn}

\author{Haoyu Wang}
\affiliation{%
  \institution{Institute of Microelectronics, University of the Chinese Academy of Sciences}
  \city{Beijing}
  \country{China}
}
\email{wanghaoyu252@mails.ucas.ac.cn}

\author{Shun Huang}
\affiliation{%
  \institution{National Institute of Health Data Science, Peking University}
  \city{Beijing}
  \country{China}
}
\email{huangshun0815@gmail.com}

\author{Ting Chen}
\affiliation{%
  \institution{Department of Cardiology, Zhejiang University}
  \city{Hangzhou}
  \country{China}
}
\email{ct010151452@zju.edu.cn}

\author{Hongyan Li}
\authornotemark[1]
\authornote{Corresponding authors.}
\affiliation{%
  \institution{School of Intelligence Science and Technology, Peking University}
  \city{Beijing}
  \country{China}
}
\email{leehy@pku.edu.cn}

\author{Shenda Hong}
\authornotemark[2]
\affiliation{%
  \institution{National Institute of Health Data Science, Peking University}
  \city{Beijing}
  \country{China}
}
\email{hongshenda@pku.edu.cn}

\renewcommand{\shortauthors}{Xiaocheng Fang et al.}

\begin{abstract}
  Cardiac Magnetic Resonance (CMR) imaging provides a comprehensive assessment of cardiac structure and function but remains constrained by high acquisition costs and reliance on expert annotations, limiting the availability of large-scale labeled datasets. In contrast, electrocardiograms (ECGs) are inexpensive, widely accessible, and offer a promising modality for conditioning the generative synthesis of cine CMR. To this end, we propose ECGFlowCMR, a novel ECG-to-CMR generative framework that integrates a Phase-Aware Masked Autoencoder (PA-MAE) and an Anatomy-Motion Disentangled Flow (AMDF) to address two fundamental challenges: (1) the cross-modal temporal mismatch between multi-beat ECG recordings and single-cycle CMR sequences, and (2) the anatomical observability gap due to the limited structural information inherent in ECGs. Extensive experiments on the UK Biobank and a proprietary clinical dataset demonstrate that ECGFlowCMR can generate realistic cine CMR sequences from ECG inputs, enabling scalable pretraining and improving performance on downstream cardiac disease classification and phenotype prediction tasks. The code is available at \url{https://github.com/PKUDigitalHealth/ECGFlowCMR}.
\end{abstract}



\maketitle

\begin{figure*}[t]
\centering
\includegraphics[width=0.95\linewidth]{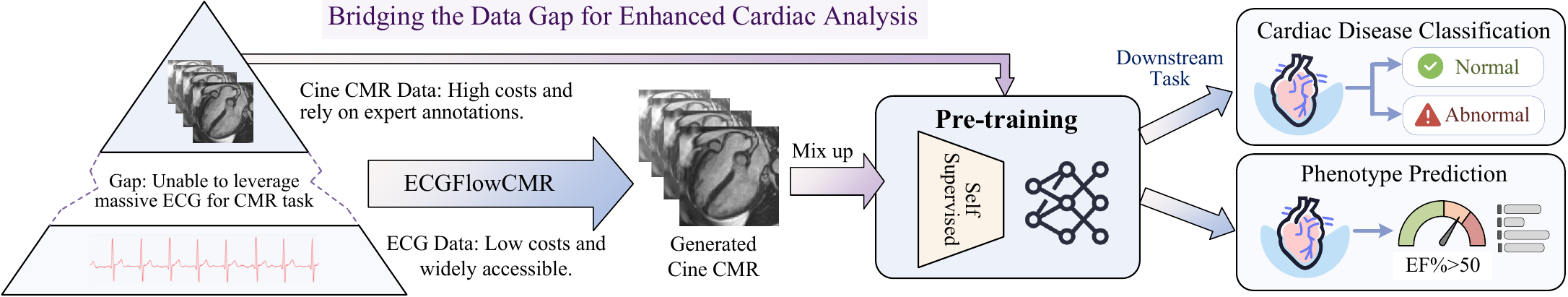}
\caption{Overview of the generative pretraining paradigm. Unlike conventional methods that depend on limited annotated CMR data, ECGFlowCMR leverages available ECG signals to synthesize realistic cine CMR sequences.}
\label{Figure6}
\end{figure*}

\section{Introduction}
Cardiac Magnetic Resonance (CMR) imaging is a pivotal noninvasive modality for the comprehensive assessment of cardiac structure and function~\cite{bai2020population,wang2024screening}. However, the development of AI models for automated CMR analysis is hindered by the scarcity of large-scale, expertly annotated datasets. Although the UK Biobank (UKB)~\cite{bycroft2018uk} provides 42,129 CMR scans, this scale remains significantly below the sample size (typically on the order of $10^5$ to $10^6$) required to pretrain robust foundation models~\cite{abbaspourazadlarge,moor2023foundation,nie2025anyppg}. The substantial cost and annotation burden associated with CMR acquisition have so far precluded the release of larger public datasets, thereby limiting the generalizability and clinical utility of AI-based methods. In this context, generative modeling has emerged as a promising approach to synthesize realistic CMR data at scale, mitigating data scarcity and enhancing model performance.

Recent progress in generative modeling has enabled high-fidelity synthesis across various medical imaging and physiological modalities, such as chest X-rays~\cite{bluethgen2025vision,prakash2025evaluating}, 3D brain MRIs~\cite{peng2023generating,tudosiu2024realistic}, ECG waveforms~\cite{fang2025ppgflowecg,wang2026se}, and echocardiographic videos~\cite{chen2024ultrasound,liechopulse}, thereby improving representation learning and downstream task performance. However, cine CMR synthesis remains comparatively underexplored. Existing methods exhibit notable limitations: DragNet~\cite{zakeri2023dragnet} relies on static-frame conditioning, which undermines temporal coherence, while CPGG~\cite{li2025phenotype} depends on phenotype labels extracted from CMR segmentation, introducing annotation dependency and limiting scalability. These issues highlight the need for flexible, annotation-independent generative frameworks for cine CMR synthesis. Given the limited availability and high acquisition cost of CMR data, synthesizing cine CMR sequences from electrocardiograms (ECGs) represents a promising and scalable alternative. ECGs are inexpensive, widely accessible, and routinely collected in both clinical practice and population-scale screening. Furthermore, ECGs capture electrophysiological signals correlated with cardiac structure and function~\cite{zhang2026ecgomics,jin2026ecg,jin2025self,bao2026position}, offering a physiologically grounded proxy for data-driven cine modeling. Preliminary work includes a cross-modal autoencoder~\cite{radhakrishnan2023cross} that learns a shared latent space for cine sequence imputation, and CardioNets~\cite{ding2026generating}, which employs masked autoregression to generate latent CMR representations. These studies demonstrate the feasibility of ECG-conditioned cine CMR synthesis and motivate the development of more expressive and generalizable models.

However, ECG-to-CMR synthesis faces two key challenges: \textbf{1) Cross-modal temporal mismatch:} A 10-second ECG captures multi-beat electrophysiological dynamics with variable heart rates, whereas a 50-frame cine CMR typically represents a single, phase-resolved cardiac cycle. This discrepancy renders temporal alignment and beat-to-cycle correspondence inherently ambiguous, complicating sequence generation. \textbf{2) Anatomical observability gap:} As ECG primarily reflects surface-level electrical activity rather than structural morphology, it imposes only weak constraints on cardiac anatomy and image appearance. This under-constrained mapping often results in synthetic CMRs that capture coarse motion patterns but fail to reconstruct fine-grained anatomical details.

To overcome these limitations, we propose ECGFlowCMR, a generative framework that integrates a Phase-Aware Masked Autoencoder (PA-MAE) and an Anatomy-Motion Disentangled Flow (AMDF) module. PA-MAE learns ECG representations under joint supervision of signal reconstruction and cardiac-phase prediction, where a dedicated phase head identifies complete cardiac cycles to enable temporal alignment with cine CMR sequences. AMDF constructs a time-invariant anatomical prior via a 3D variational autoencoder (3D-VAE), generating a static latent template as a structural anchor. Conditioned on this template and an initial motion state, a Diffusion Transformer (DiT)-based flow-matching network predicts ECG-conditioned velocity fields to synthesize anatomically accurate and temporally coherent motion. Experiments on the UK Biobank and a proprietary clinical dataset demonstrate that ECGFlowCMR generates realistic cine CMR sequences from ECG inputs, facilitates scalable pretraining, and enhances performance on downstream cardiac analysis tasks. The main contributions of this work are summarized as follows:

\begin{figure*}[t]
\centering
\includegraphics[width=0.95\linewidth]{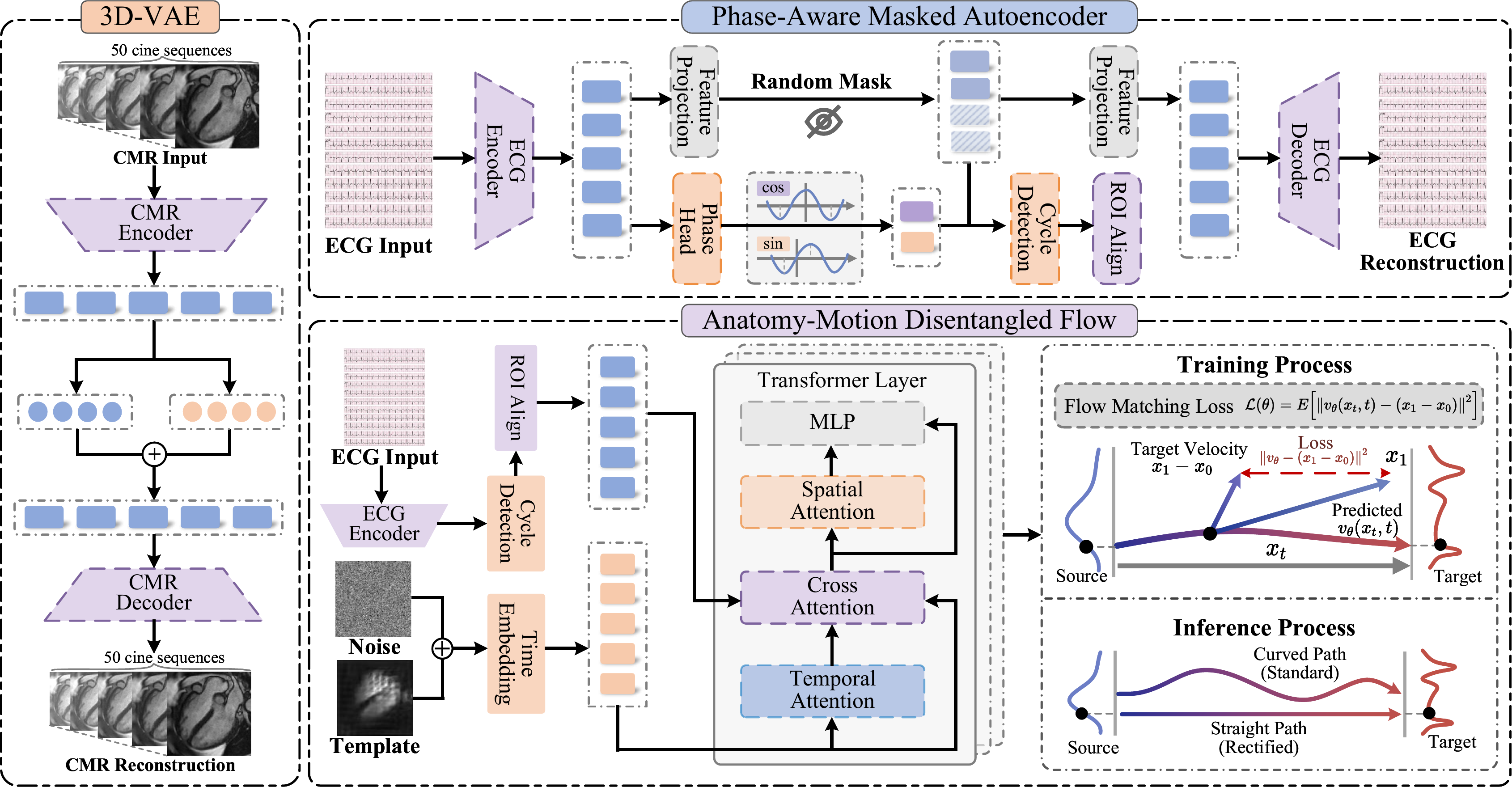}
\caption{Overview of ECGFlowCMR for generating realistic cine CMR sequences from 12-lead ECGs. The framework comprises two key components: PA-MAE and AMDF.}
\label{Figure1}
\end{figure*}

\begin{itemize}
\item We propose ECGFlowCMR, a novel ECG-to-CMR generative framework designed for scalable pretraining and downstream cardiac analysis tasks. It addresses two major limitations of prior approaches: the cross-modal temporal mismatch and the anatomical observability gap.
\item The proposed PA-MAE module enables cycle-level alignment between multi-beat ECG signals and single-cycle cine CMR by modeling cardiac phase dynamics, while the AMDF module improves anatomical plausibility by disentangling static structure from dynamic motion in the latent space.
\item Extensive experiments on the UK Biobank dataset demonstrate that ECGFlowCMR synthesizes high-fidelity cine CMR and improves downstream disease classification and phenotype prediction when used for synthetic pretraining.
\item To evaluate generalizability, we further curate a proprietary clinical dataset of 535 patients for cardiomyopathy classification. Results on this external benchmark confirm our method’s robustness under real-world distribution shifts in patient demographics and clinical acquisition settings.
\end{itemize}

\section{Related Work}
\subsection{Video Generation}
Video generation has evolved from direct pixel-level synthesis to scalable latent sequence modeling, aiming to preserve spatial fidelity and temporal coherence. Early GAN-based models, such as VGAN~\cite{vondrick2016generating}, TGAN~\cite{saito2017temporal}, and MoCoGAN~\cite{tulyakov2018mocogan}, captured short-term dynamics but struggled with long-range consistency and scalability in high-dimensional spatiotemporal spaces. To reduce visual redundancy and improve temporal modeling efficiency, many methods generate videos in compressed latent spaces. Stochastic latent-variable models further alleviate pixel-level complexity and support long-horizon prediction~\cite{babaeizadeh2017stochastic,denton2018stochastic}. Discrete token-based approaches encode frames or spatiotemporal patches with vector-quantized representations, enabling Transformer-based autoregression, masked modeling, and hierarchical planning~\cite{van2017neural,yan2021videogpt,villegas2022phenaki}. More recently, diffusion models have achieved strong perceptual quality through iterative denoising with temporal coupling~\cite{ho2022video}, and latent diffusion variants further improve sample fidelity and computational efficiency~\cite{he2022latent,blattmann2023align}. Overall, recent video generation models increasingly combine latent compression with expressive temporal backbones, including Transformers and diffusion models, while incorporating conditioning signals such as text or actions to enhance controllability and long-range temporal consistency~\cite{singer2022make,ho2022video,villegas2022phenaki}.

\subsection{CMR Video Generation}
Cine CMR generation synthesizes temporally coherent cardiac sequences with accurate anatomy and physiologically plausible motion. Early methods model motion by warping a reference frame with estimated deformation fields, as in DragNet~\cite{zakeri2023dragnet}, which couples spatiotemporal registration and generation. Recent models learn cardiac video distributions with anatomical or functional priors. GANcMRI~\cite{vukadinovic2023gancmri} controls morphology and function via latent prompting, while diffusion models improve fidelity and motion realism through disease-aware 4D synthesis~\cite{liu2024texdc} and latent or patch-wise efficient designs~\cite{lei2025patch}. Intra-CMR translation transfers motion and anatomical priors across acquisition protocols under weak or unpaired supervision, including tagged-to-cine Transformers~\cite{liu2024tagged} and motion-guided DENSE-to-cine diffusion~\cite{deb2025unsupervised}. These methods require sequence-specific CMR inputs at inference, limiting large-scale synthesis. Phenotype-conditioned generators, such as CPGG~\cite{li2025phenotype}, reduce this requirement by using structured cardiac phenotypes but depend on reliable phenotype extraction. ECG-conditioned frameworks further synthesize cine CMR from ECG signals by aligning ECG and CMR latents~\cite{radhakrishnan2023cross} or using masked autoregressive decoding, as in CardioNets~\cite{ding2026generating}. Yet ECG-conditioned generation remains constrained by cross-modal temporal misalignment and incomplete anatomical observability.

\section{Methodology}
\subsection{Overview of ECGFlowCMR}
As illustrated in Figure~\ref{Figure1}, ECGFlowCMR generates cine CMR sequences from 12-lead ECGs through two coupled modules. PA-MAE learns rhythm-aware ECG representations by combining masked signal reconstruction with cardiac-phase prediction; the predicted phases identify complete cardiac cycles, which are resampled by ROI Align into 50-frame conditioning representations aligned with cine CMR. AMDF then synthesizes cine CMR videos in a compact latent space by constructing a 3D-VAE anatomical template and learning ECG-conditioned latent velocity fields with a DiT-based flow-matching network. This phase-aligned conditioning and anatomy-motion disentangled synthesis support anatomically plausible and temporally coherent ECG-to-CMR generation for scalable generative pretraining.

\subsection{Phase-Aware Masked Autoencoder}
To extract semantically rich and rhythm-aware representations from 12-lead ECGs, we propose the Phase-Aware Masked Autoencoder (PA-MAE). Trained with dual supervision, PA-MAE facilitates robust morphological representation learning and enables precise temporal alignment with cine CMR sequences.

\paragraph{{\bfseries Masked Signal Reconstruction.}}
As illustrated in Figure~\ref{Figure1}, PA-MAE adopts a standard masked autoencoder~\cite{he2022masked} for self-supervised ECG representation learning. Given an input $x_\text{ecg} \in \mathbb{R}^{C \times T}$, where $C$ denotes the number of leads and $T$ the temporal length, the encoder $\mathbf{E}_{\text{ecg}}$ extracts semantic features as $F = \mathbf{E}_{\text{ecg}}(x_\text{ecg})$. To promote robust feature learning, a proportion $\rho$ of temporal positions is randomly masked by zeroing out the corresponding feature vectors:
\begin{equation}
F_{\text{masked}} = F \odot m,
\end{equation}
where $m \in \{0,1\}^{T'} $ is a binary mask containing a fraction $\rho$ of zeros, $T'$ denotes the encoded temporal length, and $\odot$ represents element-wise multiplication. The masked features $F_{\text{masked}}$ are then passed through a decoder $\mathbf{D}_{\text{ecg}}$ to reconstruct the original ECG signal. Specifically, the reconstruction is given by $\hat{x}_{\text{ecg}} = \mathbf{D}_{\text{ecg}}(F_{\text{masked}})$, where $\hat{x}_{\text{ecg}}$ denotes the predicted signal corresponding to the input. The model is trained to minimize the mean squared error (MSE) between the input and the reconstructed signal:
\begin{equation}
\mathcal{L}_{\text{rec}} = \frac{1}{CT}\left\lVert \hat{x}_{\text{ecg}} - x_{\text{ecg}} \right\rVert_2^2,
\end{equation}
which encourages the model to capture meaningful morphological patterns without requiring manual annotations.

\paragraph{{\bfseries Phase-Aware Temporal Supervision.}} 
Beyond masked signal reconstruction, we propose a dedicated phase prediction head $\mathbf{P}_{\text{ecg}}$ to explicitly model cardiac rhythm and mitigate temporal misalignment between ECG signals and CMR sequences. The module $\mathbf{P}_{\text{ecg}}$ operates on the full set of unmasked encoder features and outputs sinusoidal representations of the cardiac phase:
\begin{equation}
\hat{\phi} = \mathbf{P}_{\text{ecg}}(F) = [\sin(\phi), \cos(\phi)],
\end{equation}
where $\phi \in [0, 2\pi]$ denotes the normalized phase within each cardiac cycle. Ground-truth phase labels $\phi_{\text{gt}}$ are computed via R-peak detection from Lead II and linearly interpolated across each R-R interval. The phase prediction loss is followed:
\begin{equation}
\mathcal{L}_{\text{phase}} = \frac{1}{T'} \|\hat{\phi} - \phi_{\text{gt}}\|_2^2.
\end{equation}
Here, $T'$ denotes the encoded temporal length. To achieve temporal alignment between ECG and cine CMR, complete cardiac cycles are extracted from the predicted phase sequence and resampled using ROI Align into 50-frame representations, matching the temporal resolution of cine CMR.

\paragraph{{\bfseries Training Objective.}} The overall training objective combines both reconstruction and phase prediction losses:
\begin{equation}
\mathcal{L}_{\text{PA-MAE}} = \mathcal{L}_{\text{rec}} + \mathcal{L}_{\text{phase}}.
\end{equation}
This dual-supervision strategy ensures that the learned features retain both morphological and temporal semantics critical for downstream CMR synthesis.

\subsection{Anatomy-Motion Disentangled Flow}
To address the anatomical observability gap where ECG signals impose weak constraints on cardiac morphology, we propose the Anatomy-Motion Disentangled Flow (AMDF), which explicitly disentangles static anatomical structure from dynamic temporal motion in a latent space. Specifically, AMDF first extracts a time-invariant anatomical representation using a 3D variational autoencoder (3D-VAE), and subsequently models the time-dependent motion residual via a conditional flow-matching network.

\paragraph{{\bfseries Anatomical Anchor Learning.}}
We employ a 3D-VAE to encode CMR videos into a latent space using an anisotropic compression strategy that reduces spatial resolution while preserving temporal fidelity. Given a CMR video $x_\text{cmr} \in \mathbb{R}^{C \times T \times H \times W}$, where $C$ denotes the input channel and $T$ the number of frames, the encoder $\mathbf{E}_{\text{cmr}}$ maps $x_\text{cmr}$ to a latent representation $\mathbf{z} = \mathbf{E}_{\text{cmr}}(x_\text{cmr})$, where $\mathbf{z} \in \mathbb{R}^{C' \times T \times H' \times W'}$, with $C'$ indicating the latent channel dimension and $H' \times W'$ the spatially downsampled resolution. The decoder $\mathbf{D}_{\text{cmr}}$ then reconstructs the input as $\hat{x}_\text{cmr} = \mathbf{D}_{\text{cmr}}(\mathbf{z})$.

After training the 3D-VAE, we extract latent representations from all training CMR videos and compute a static anatomical template by averaging over both the sample and temporal dimensions:
\begin{equation}
\mathbf{z}_{\text{template}} = 
\frac{1}{N T} 
\sum_{i=1}^{N} 
\sum_{t=1}^{T} 
\mathbf{z}_{i,t},
\end{equation}
where $N$ denotes the number of training subjects and $T$ the number of frames per sample. The resulting template representation $\mathbf{z}_{\text{template}}$ captures the average anatomical structure across the training population and serves as a structural prior encoding population-level cardiac morphology.

\begin{table*}[t]
\centering
\small
\renewcommand\tabcolsep{12pt}
\renewcommand\arraystretch{1.0}
\newcommand{\red}[1]{\textcolor{red}{#1}}
\caption{Quantitative comparison of cine CMR generation quality and inference efficiency on the UKB dataset.}
\begin{tabular}{lcccccc}
\toprule
\textbf{Methods} & \textbf{Ref.} & \textbf{LPIPS$\downarrow$} & \textbf{FID$\downarrow$} & \textbf{FVD$\downarrow$} & \textbf{Inference time$\downarrow$} \\
\midrule
VideoGPT~\cite{yan2021videogpt} & ArXiv'21 & 0.35 & 107.41 & 24.99 & 2.15 sec / vid. \\

ModelScopeT2V~\cite{wang2023modelscope} & ArXiv'23 & 0.37 & 104.19 & 31.43 & 3.87 sec / vid. \\

Cross-Modal Autoencoder~\cite{radhakrishnan2023cross} & Nat. Commun.'23 & 0.32 & 99.24 & 28.62 & 0.63 sec / vid. \\

EchoPulse~\cite{liechopulse} & ICLR'25 & 0.31 & 85.80  & 21.41 & 0.84 sec / vid. \\

CardioNets~\cite{ding2026generating} & NEJM AI'26 & 0.28 & 89.83  & 21.65 & 4.33 sec / vid. \\

\midrule
\rowcolor{gray!20}
ECGFlowCMR & Ours & \textbf{0.27} & \textbf{37.28} & \textbf{14.41} & \textbf{0.45 sec / vid.} \\
\rowcolor{gray!20}
Increased & - & \red{3.57\%$\uparrow$} & \red{56.56\%$\uparrow$} & \red{32.70\%$\uparrow$} & \red{28.57\%$\uparrow$} \\
\bottomrule
\end{tabular}
\label{Table1}
\end{table*}

\paragraph{{\bfseries Conditional Flow Matching.}}
To model time-varying cardiac motion in a physiologically coherent manner, we adopt a conditional flow-matching framework built upon a Diffusion Transformer (DiT)~\cite{peebles2023scalable,liuflow}. Conditioned on a static anatomical template and an initial motion state, the network learns a continuous velocity field that governs latent-space dynamics. Specifically, we define a velocity field $v_\theta(\mathbf{z}_t, t, c)$ that predicts the instantaneous motion at an intermediate time $t \in [0, 1]$, where $\mathbf{z}_t$ denotes the interpolated latent representation at time $t$, and $c$ represents ECG-derived conditioning features extracted by the PA-MAE encoder.

During training, we simulate latent motion trajectories by sampling time points $t \sim \mathbf{U}(0, 1)$ and linearly interpolating between a noise-perturbed anatomical anchor $\mathbf{z}_0$ and the target CMR latent representation $\mathbf{z}_1$:
\begin{equation}
\mathbf{z}_t = (1 - t) \cdot \mathbf{z}_0 + t \cdot \mathbf{z}_1,
\end{equation}
where the noise-perturbed anatomical anchor $\mathbf{z}_0$ is defined as:
\begin{equation}
\mathbf{z}_0 = \mathbf{z}_{\text{template}} + \alpha \cdot \boldsymbol{\epsilon}, \quad \boldsymbol{\epsilon} \sim \mathcal{N}(0, \mathbf{I}).
\label{eq8}
\end{equation}
Here, $\mathbf{z}_1$ is the latent representation of the target CMR sequence obtained from the 3D-VAE encoder, and $\alpha$ controls the level of noise injection to the anatomical prior.

The ground-truth velocity field $\mathbf{v}_{\text{true}} = \mathbf{z}_1 - \mathbf{z}_0$ defines a constant linear drift from the initial to the target state. The DiT-based network $v_\theta$ is trained to regress this velocity, conditioned on the interpolated latent $\mathbf{z}_t$, the corresponding time step $t$, and the ECG-derived context $c$:
\begin{equation}
\hat{\mathbf{v}} = v_\theta(\mathbf{z}_t, t, c).
\end{equation}
The training objective minimizes the expected mean squared error (MSE) between the predicted and ground-truth velocity fields:
\begin{equation}
\mathcal{L}_{\text{AMDF}} = \mathbb{E}_{t, \mathbf{z}_0, \mathbf{z}_1, c} \left[ \|v_\theta(\mathbf{z}_t, t, c) - (\mathbf{z}_1 - \mathbf{z}_0)\|_2^2 \right].
\end{equation}
This objective encourages the network to learn phase-aware latent trajectories that maintain anatomical consistency and temporal coherence, thereby enabling realistic cardiac motion synthesis.

\paragraph{{\bfseries Inference and Sampling.}}
At inference time, we synthesize cine CMR latents by numerically integrating the learned velocity field from an initial latent state $\mathbf{z}_0$ toward the target latent $\mathbf{z}_1$. Starting from the anatomical template perturbed with noise, we iteratively update the latent using an explicit Euler scheme:
\begin{equation}
\mathbf{z}_{t+\Delta t} = \mathbf{z}_t + \Delta t \cdot v_\theta(\mathbf{z}_t, t, c).
\end{equation}
Here, $\Delta t$ denotes the integration step size and $v_\theta$ is the DiT-based conditional velocity field. The integration is performed from $t = 0$ to $t = 1$, yielding the final latent representation $\mathbf{z}_1$. This latent is then decoded by the 3D-VAE decoder to generate the corresponding cine CMR sequence.

By jointly refining anatomical structure via the anatomical anchor and synthesizing ECG-guided temporal motion through conditional flow matching, AMDF generates cardiac dynamics that are anatomically plausible and temporally coherent, effectively mitigating the anatomical observability gap between ECG signals and cardiac morphology.

\begin{table*}[t]
    \small
    \centering
    \renewcommand\tabcolsep{10pt}
    \renewcommand\arraystretch{1.0}

    \newcommand{\res}[2]{$#1_{\pm #2}$}
    \newcommand{\bestres}[2]{\bm{$#1_{\pm #2}$}}
    \newcommand{\secondres}[2]{\textcolor{red}{\bm{$#1_{\pm #2}$}}}

    \caption{Cardiac disease classification performance on UKB cohorts under different synthetic-data mixing ratios.}
    \begin{tabular}{lccccccc}
        \toprule
        \multirow{2}{*}{\textbf{Methods}} & \multirow{2}{*}{\textbf{Ref.}} &
        \multicolumn{2}{c}{\textbf{UKB-CAD}} & \multicolumn{2}{c}{\textbf{UKB-CM}} & \multicolumn{2}{c}{\textbf{UKB-HF}} \\
        \cmidrule(lr){3-4} \cmidrule(lr){5-6} \cmidrule(lr){7-8}
        & & \textbf{ACC$\uparrow$} & \textbf{AUC$\uparrow$} & \textbf{ACC$\uparrow$} & \textbf{AUC$\uparrow$} & \textbf{ACC$\uparrow$} & \textbf{AUC$\uparrow$} \\
        \midrule
        ViT~\cite{dosovitskiy2020image} & ICLR'21
        & \res{0.662}{0.014} & \res{0.714}{0.022} & \res{0.705}{0.008} & \res{0.747}{0.032} & \res{0.708}{0.038} & \res{0.779}{0.018} \\
        MAE~\cite{he2022masked}(real) & CVPR'22
        & \res{0.695}{0.086} & \res{0.743}{0.048} & \res{0.776}{0.017} & \res{0.806}{0.054} & \res{0.776}{0.032} & \res{0.854}{0.016} \\

        \midrule
        \multicolumn{8}{l}{\textit{Training with 100\% Mixed Synthetic Data}} \\
        VideoGPT~\cite{yan2021videogpt} & ArXiv'21
        & \res{0.697}{0.045} & \res{0.752}{0.021} & \res{0.779}{0.051} & \res{0.809}{0.038} & \res{0.781}{0.025} & \res{0.857}{0.012} \\
        ModelScopeT2V~\cite{wang2023modelscope} & ArXiv'23
        & \res{0.701}{0.031} & \res{0.761}{0.052} & \res{0.784}{0.024} & \res{0.811}{0.042} & \res{0.784}{0.054} & \res{0.859}{0.031} \\
        EchoPulse~\cite{liechopulse} & ICLR'25
        & \res{0.709}{0.012} & \res{0.778}{0.035} & \res{0.799}{0.007} & \res{0.828}{0.048} & \res{0.798}{0.032} & \res{0.868}{0.019} \\

        CardioNets~\cite{ding2026generating} & NEJM AI'26
        & \res{0.706}{0.054} & \res{0.769}{0.018} & \res{0.792}{0.037} & \res{0.819}{0.014} & \res{0.791}{0.009} & \res{0.861}{0.045} \\

        \rowcolor{gray!20}
        ECGFlowCMR & Ours
        & \bestres{0.716}{0.027} & \bestres{0.787}{0.011} & \bestres{0.806}{0.035} & \bestres{0.837}{0.024} & \bestres{0.808}{0.018} & \bestres{0.876}{0.026} \\
        \rowcolor{gray!20}
        Increased & -
        & \textcolor{red}{0.99\%$\uparrow$} 
        & \textcolor{red}{1.16\%$\uparrow$} 
        & \textcolor{red}{0.88\%$\uparrow$} 
        & \textcolor{red}{1.09\%$\uparrow$} 
        & \textcolor{red}{1.25\%$\uparrow$} 
        & \textcolor{red}{0.92\%$\uparrow$} \\

        \midrule
        \multicolumn{8}{l}{\textit{Training with 200\% Mixed Synthetic Data}} \\
        VideoGPT~\cite{yan2021videogpt} & ArXiv'21
        & \res{0.699}{0.028} & \res{0.765}{0.044} & \res{0.788}{0.019} & \res{0.815}{0.052} & \res{0.789}{0.041} & \res{0.856}{0.025} \\
        ModelScopeT2V~\cite{wang2023modelscope} & ArXiv'23
        & \res{0.705}{0.051} & \res{0.774}{0.015} & \res{0.796}{0.044} & \res{0.823}{0.031} & \res{0.797}{0.012} & \res{0.864}{0.053} \\
        EchoPulse~\cite{liechopulse} & ICLR'25
        & \res{0.717}{0.014} & \res{0.789}{0.058} & \res{0.810}{0.053} & \res{0.839}{0.029} & \res{0.811}{0.026} & \res{0.878}{0.014} \\

        CardioNets~\cite{ding2026generating} & NEJM AI'26
        & \res{0.711}{0.036} & \res{0.782}{0.027} & \res{0.804}{0.022} & \res{0.831}{0.015} & \res{0.804}{0.048} & \res{0.871}{0.037} \\
        \rowcolor{gray!20}
        ECGFlowCMR & Ours
        & \bestres{0.723}{0.019} & \bestres{0.796}{0.028} & \bestres{0.818}{0.013} & \bestres{0.845}{0.039} & \bestres{0.817}{0.030} & \bestres{0.886}{0.021} \\
        \rowcolor{gray!20}
        Increased & -
        & \textcolor{red}{0.84\%$\uparrow$} 
        & \textcolor{red}{0.89\%$\uparrow$} 
        & \textcolor{red}{0.99\%$\uparrow$} 
        & \textcolor{red}{0.72\%$\uparrow$} 
        & \textcolor{red}{0.74\%$\uparrow$} 
        & \textcolor{red}{0.91\%$\uparrow$} \\

        \midrule
        \multicolumn{8}{l}{\textit{Training with 300\% Mixed Synthetic Data}} \\
        VideoGPT~\cite{yan2021videogpt} & ArXiv'21
        & \res{0.709}{0.034} & \res{0.776}{0.051} & \res{0.798}{0.042} & \res{0.825}{0.016} & \res{0.800}{0.054} & \res{0.863}{0.032} \\
        ModelScopeT2V~\cite{wang2023modelscope} & ArXiv'23
        & \res{0.714}{0.047} & \res{0.783}{0.023} & \res{0.805}{0.014} & \res{0.833}{0.051} & \res{0.807}{0.021} & \res{0.870}{0.044} \\
        EchoPulse~\cite{liechopulse} & ICLR'25
        & \res{0.723}{0.052} & \res{0.797}{0.013} & \res{0.818}{0.028} & \res{0.848}{0.054} & \res{0.820}{0.012} & \res{0.884}{0.036} \\

        CardioNets~\cite{ding2026generating} & NEJM AI'26
        & \res{0.718}{0.025} & \res{0.790}{0.048} & \res{0.812}{0.033} & \res{0.841}{0.027} & \res{0.813}{0.045} & \res{0.877}{0.019} \\

        \rowcolor{gray!20}
        ECGFlowCMR & Ours
        & \bestres{0.730}{0.011} & \bestres{0.804}{0.033} & \bestres{0.826}{0.020} & \bestres{0.854}{0.014} & \bestres{0.826}{0.028} & \bestres{0.891}{0.016} \\
        \rowcolor{gray!20}
        Increased & -
        & \textcolor{red}{0.97\%$\uparrow$} 
        & \textcolor{red}{0.88\%$\uparrow$} 
        & \textcolor{red}{0.98\%$\uparrow$} 
        & \textcolor{red}{0.71\%$\uparrow$} 
        & \textcolor{red}{0.73\%$\uparrow$} 
        & \textcolor{red}{0.79\%$\uparrow$} \\

        \bottomrule
    \end{tabular}
    \label{Table2}
\end{table*}

\begin{table*}[t]
\small
\centering
\renewcommand\tabcolsep{8pt}
\renewcommand\arraystretch{1.0}

\newcommand{\red}[1]{\textcolor{red}{#1}}

\caption{Cardiac phenotype prediction performance on the UKB dataset under different synthetic-data mixing ratios, with Overall denoting the average $R^2$ across 82 cardiac phenotypes.}
\begin{tabular}{lcccccccccc}
\toprule
\multirow{2}{*}{\textbf{Methods}} & \multirow{2}{*}{\textbf{Ref.}} &
\multicolumn{2}{c}{\textbf{LVEDV}} & \multicolumn{2}{c}{\textbf{LVEF}} &
\multicolumn{2}{c}{\textbf{LVM}} & \multicolumn{2}{c}{\textbf{RVEDV}} & \textbf{Overall} \\
\cmidrule(lr){3-4} \cmidrule(lr){5-6} \cmidrule(lr){7-8} \cmidrule(lr){9-10} \cmidrule(lr){11-11}
 &  & \textbf{MAE$\downarrow$} & $\mathbf{R^{2}\uparrow}$
 & \textbf{MAE$\downarrow$} & $\mathbf{R^{2}\uparrow}$
 & \textbf{MAE$\downarrow$} & $\mathbf{R^{2}\uparrow}$
 & \textbf{MAE$\downarrow$} & $\mathbf{R^{2}\uparrow}$ & $\mathbf{R^{2}\uparrow}$ \\
\midrule
ViT~\cite{dosovitskiy2020image} & ICLR'21 & 11.52 & 0.772 & 3.75 & 0.367 & 7.56 & 0.763 & 13.14 & 0.770 & 0.417 \\
MAE~\cite{he2022masked} (real) & CVPR'22 & 11.35 & 0.769 & 3.67 & 0.412 & 6.67 & 0.801 & 12.62 & 0.784 & 0.441 \\
\midrule

\multicolumn{11}{l}{\textit{Training with 100\% Mixed Synthetic Data}} \\
VideoGPT~\cite{yan2021videogpt} & ArXiv'21 & 11.20 & 0.773 & 3.54 & 0.415 & 6.48 & 0.805 & 12.47 & 0.786 & 0.445 \\
ModelScopeT2V~\cite{wang2023modelscope} & ArXiv'23 & 11.28 & 0.771 & 3.51 & 0.418 & 6.51 & 0.808 & 12.40 & 0.790 & 0.443 \\
EchoPulse~\cite{liechopulse} & ICLR'25
& 10.83 & 0.785 & 3.44 & 0.427 & 6.25 & 0.818 & 12.16 & 0.802 & 0.454 \\
CardioNets~\cite{ding2026generating} & NEJM AI'26 & 10.97 & 0.782 & 3.47 & 0.423 & 6.32 & 0.814 & 12.25 & 0.799 & 0.451 \\
\rowcolor{gray!20}
ECGFlowCMR & Ours
& \textbf{10.40} & \textbf{0.806} & \textbf{3.41} & \textbf{0.436} & \textbf{6.14} & \textbf{0.828} & \textbf{11.92} & \textbf{0.809} & \textbf{0.470} \\
\rowcolor{gray!20}
Increased & -
& \textcolor{red}{3.97\%$\uparrow$} 
& \textcolor{red}{2.68\%$\uparrow$} 
& \textcolor{red}{0.87\%$\uparrow$} 
& \textcolor{red}{2.11\%$\uparrow$} 
& \textcolor{red}{1.76\%$\uparrow$} 
& \textcolor{red}{1.22\%$\uparrow$} 
& \textcolor{red}{1.97\%$\uparrow$} 
& \textcolor{red}{0.87\%$\uparrow$} 
& \textcolor{red}{3.52\%$\uparrow$} \\
\midrule

\multicolumn{11}{l}{\textit{Training with 200\% Mixed Synthetic Data}} \\
VideoGPT~\cite{yan2021videogpt} & ArXiv'21 & 11.24 & 0.774 & 3.51 & 0.417 & 6.44 & 0.806 & 12.49 & 0.789 & 0.447 \\
ModelScopeT2V~\cite{wang2023modelscope} & ArXiv'23 & 11.25 & 0.773 & 3.45 & 0.421 & 6.35 & 0.812 & 12.27 & 0.795 & 0.452 \\
EchoPulse~\cite{liechopulse} & ICLR'25
& 10.69 & 0.795 & 3.40 & 0.433 & 6.14 & 0.821 & 11.89 & 0.806 & 0.463 \\
CardioNets~\cite{ding2026generating} & NEJM AI'26 & 10.82 & 0.791 & 3.43 & 0.429 & 6.24 & 0.817 & 12.04 & 0.803 & 0.459 \\
\rowcolor{gray!20}
ECGFlowCMR & Ours
& \textbf{10.19} & \textbf{0.812} & \textbf{3.39} & \textbf{0.441} & \textbf{5.92} & \textbf{0.836} & \textbf{11.53} & \textbf{0.821} & \textbf{0.482} \\
\rowcolor{gray!20}
Increased & -
& \textcolor{red}{4.68\%$\uparrow$} 
& \textcolor{red}{2.14\%$\uparrow$} 
& \textcolor{red}{0.29\%$\uparrow$} 
& \textcolor{red}{1.85\%$\uparrow$} 
& \textcolor{red}{3.58\%$\uparrow$} 
& \textcolor{red}{1.83\%$\uparrow$} 
& \textcolor{red}{3.03\%$\uparrow$} 
& \textcolor{red}{1.86\%$\uparrow$} 
& \textcolor{red}{4.10\%$\uparrow$} \\
\midrule

\multicolumn{11}{l}{\textit{Training with 300\% Mixed Synthetic Data}} \\
VideoGPT~\cite{yan2021videogpt} & ArXiv'21 & 11.13 & 0.778 & 3.47 & 0.420 & 6.39 & 0.809 & 12.42 & 0.791 & 0.450 \\
ModelScopeT2V~\cite{wang2023modelscope} & ArXiv'23 & 11.18 & 0.775 & 3.42 & 0.425 & 6.31 & 0.815 & 12.20 & 0.797 & 0.456 \\
EchoPulse~\cite{liechopulse} & ICLR'25
& 10.44 & 0.804 & 3.36 & 0.435 & 6.04 & 0.826 & 11.79 & 0.811 & 0.471 \\
CardioNets~\cite{ding2026generating} & NEJM AI'26 & 10.65 & 0.798 & 3.38 & 0.434 & 6.17 & 0.821 & 11.84 & 0.808 & 0.465 \\
\rowcolor{gray!20}
ECGFlowCMR & Ours
& \textbf{9.97} & \textbf{0.821} & \textbf{3.32} & \textbf{0.442} & \textbf{5.84} & \textbf{0.844} & \textbf{11.32} & \textbf{0.834} & \textbf{0.499} \\
\rowcolor{gray!20}
Increased & -
& \textcolor{red}{4.50\%$\uparrow$} 
& \textcolor{red}{2.11\%$\uparrow$} 
& \textcolor{red}{1.19\%$\uparrow$} 
& \textcolor{red}{1.61\%$\uparrow$} 
& \textcolor{red}{3.31\%$\uparrow$} 
& \textcolor{red}{2.18\%$\uparrow$} 
& \textcolor{red}{3.99\%$\uparrow$} 
& \textcolor{red}{2.84\%$\uparrow$} 
& \textcolor{red}{5.94\%$\uparrow$} \\
\bottomrule
\end{tabular}
\label{Table3}
\end{table*}

\section{Experiments}
\subsection{Datasets and Evaluation Metrics}
We conducted experiments on two datasets: a large-scale public dataset from the UK Biobank (UKB) and a proprietary clinical dataset curated for cardiomyopathy diagnosis from the Affiliated Hospital of Zhejiang University, referred to as ZJU-CM. These datasets were used to evaluate both in-distribution performance and out-of-distribution generalization. All classification tasks were assessed using five-fold cross-validation. Evaluation metrics included LPIPS~\cite{zhang2018unreasonable}, FID~\cite{heusel2017gans}, FVD~\cite{unterthiner2019fvd}, and inference time.

\paragraph{{\bfseries UKB Dataset.}}
We evaluated our ECGFlowCMR on 42,129 four-chamber cine CMR sequences from the UKB, each paired with a corresponding 12-lead ECG. To avoid subject overlap, the data were split at the patient level into 29,490 training, 4,212 validation, and 8,427 testing samples. Two downstream tasks were defined. For cardiac disease classification, we constructed three balanced cohorts with equal numbers of positive and negative cases: coronary artery disease (UKB-CAD, n = 5,464), cardiomyopathy (UKB-CM, n = 196), and heart failure (UKB-HF, n = 578). For cardiac phenotype prediction, we further subset the above split into 19,487 training, 2,758 validation, and 5,553 testing samples, each annotated with 82 phenotypes following the UKB imaging protocol~\cite{bai2020population}. This dual-task setup facilitates comprehensive evaluation of both diagnostic classification and quantitative phenotype regression, aligning with the practical demands of automated cardiac image interpretation.

\paragraph{{\bfseries ZJU-CM Dataset.} } 
The ZJU-CM dataset was collected from the Affiliated Hospital of Zhejiang University and consists of 535 patients. It was curated for cardiomyopathy classification and comprises four diagnostic groups: 195 patients with hypertrophic cardiomyopathy (HCM), 160 with dilated cardiomyopathy (DCM), 33 with restrictive cardiomyopathy (RCM), and 147 healthy controls. ZJU-CM was used to evaluate two classification tasks: (i) binary classification of cardiomyopathy versus control, and (ii) multi-class classification of cardiomyopathy subtypes.

\subsection{Implementation Details and Baselines}
For data preprocessing, we utilized the UKB segmentation model~\cite{bai2020population} to extract cardiac regions and uniformly resized all cine CMR sequences to a spatial resolution of $1 \times 50 \times 96 \times 96$. A 3D-VAE was trained with an $8\times$ spatial compression ratio to encode CMR videos. Simultaneously, our PA-MAE processed 12-lead ECG signals comprising 5,000 timepoints, employing $8\times$ temporal downsampling and a random masking ratio of 0.5. The AMDF module was trained for 10 epochs using the AdamW optimizer with a learning rate of $1 \times 10^{-4}$, weight decay of $1 \times 10^{-4}$, and batch size of 4. Downstream diagnostic models were fine-tuned for 100 epochs. All experiments were conducted on an NVIDIA A100 GPU.

We compared our method with four state-of-the-art baselines: VideoGPT~\cite{yan2021videogpt}, a vector-quantized autoregressive model for video generation; ModelScopeT2V~\cite{wang2023modelscope}, a diffusion-based framework for text-to-video synthesis; EchoPulse~\cite{liechopulse}, a discrete token–based masked modeling framework for ECG-to-echocardiogram generation; and CardioNets~\cite{ding2026generating}, a masked autoregressive framework for ECG-to-CMR translation. We additionally included a cross-modal autoencoder~\cite{radhakrishnan2023cross} that reconstructs CMR sequences from ECGs as a deterministic baseline. However, its inherent one-to-one mapping precludes the modeling of stochastic variations, making it unsuitable for generative pretraining and downstream data augmentation.

\subsection{Comparisons with State-of-The-Arts}
\paragraph{{\bfseries Evaluation on CMR Generation Quality.}} 
As shown in Table~\ref{Table1}, ECGFlowCMR surpasses five state-of-the-art baselines across all evaluation metrics on the UKB dataset, establishing its superiority in perceptual quality, distributional alignment, and computational efficiency. It achieves the lowest LPIPS (0.27), indicating enhanced perceptual similarity to the ground truth. In addition, ECGFlowCMR substantially lowers the FID to 37.28 and FVD to 14.41, yielding relative improvements of 56.56\% and 32.70\%, respectively, over the strongest baseline. Additionally, it delivers the fastest inference time of 0.45 seconds per video, which constitutes a 28.57\% reduction relative to EchoPulse and up to a 90\% reduction compared to ModelScopeT2V. While the cross-modal autoencoder achieves competitive inference speed (0.63s), its deterministic nature limits generative diversity, resulting in inferior performance on LPIPS (0.32), FID (99.24), and FVD (28.62). These findings collectively underscore the effectiveness of ECGFlowCMR in synthesizing cine CMR sequences that are both high-fidelity and efficient, making it well-suited for generative pretraining and downstream clinical applications.

\paragraph{{\bfseries Evaluation on Cardiac Disease Classification.}}
Table~\ref{Table2} reports classification performance on three UKB disease cohorts (UKB-CAD, UKB-CM, and UKB-HF) under 100\%, 200\%, and 300\% mixed synthetic-data settings. ECGFlowCMR consistently achieves the highest accuracy (ACC) and area under the ROC curve (AUC) across all disease categories and data augmentation ratios. Notably, under the 100\% synthetic-data setting, ECGFlowCMR achieves ACC/AUC scores of 0.716/0.787 on UKB-CAD, 0.806/0.837 on UKB-CM, and 0.808/0.876 on UKB-HF, surpassing the strongest baseline by a substantial margin. As the synthetic data proportion increases to 200\% and 300\%, ECGFlowCMR demonstrates monotonic improvements in classification performance. These consistent gains across all cohorts indicate that the synthetic CMR sequences generated by ECGFlowCMR are both high-quality and clinically informative, thereby enhancing downstream disease classification.

\paragraph{{\bfseries Evaluation on Cardiac Phenotype Prediction.}} 
Table~\ref{Table3} reports regression results on four representative cardiac phenotypes (LVEDV, LVEF, LVM, and RVEDV) as well as the average performance across 82 phenotypic traits under 100\%, 200\%, and 300\% synthetic-data mixing settings. ECGFlowCMR consistently achieves the lowest mean absolute error (MAE) and the highest coefficient of determination ($R^2$) across all configurations. Under the 100\% synthetic-data condition, ECGFlowCMR attains an overall $R^2$ of 0.470, exceeding the strongest baseline (EchoPulse, $R^2$ = 0.454). As the proportion of synthetic data increases, the model further improves to an $R^2$ of 0.499, yielding a relative gain of 5.94\%. These consistent and substantial improvements across diverse phenotypic traits underscore the generative fidelity of ECGFlowCMR and its effectiveness in enhancing downstream phenotype prediction.

\begin{figure}[t]
\centering
\includegraphics[width=0.99\linewidth]{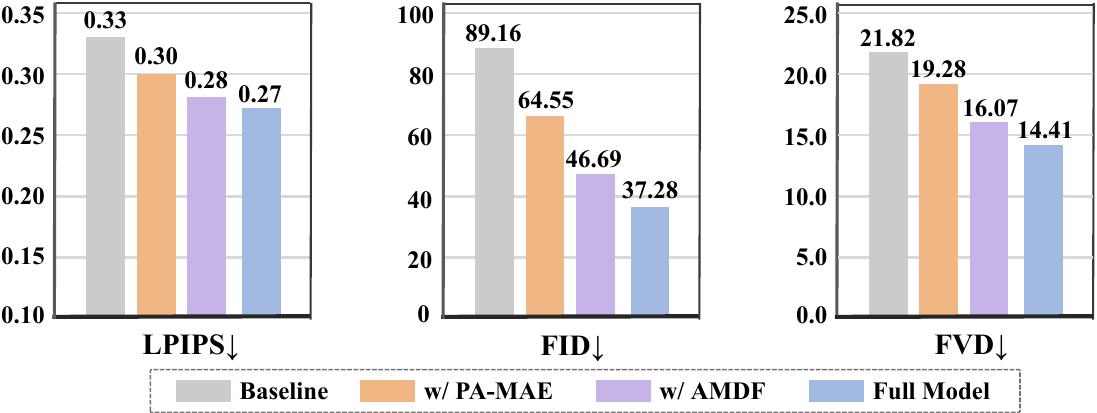}
\caption{Generation-level ablation study of PA-MAE and AMDF on the UKB dataset.}
\label{Figure2}
\end{figure}

\begin{table*}[t]
\small
\centering
\renewcommand\tabcolsep{8pt}
\renewcommand\arraystretch{0.93}

\newcommand{\ablres}[2]{$#1_{\pm #2}$}
\newcommand{\ablbestres}[2]{$\bm{#1_{\pm #2}}$}

\caption{Downstream-level ablation study of PA-MAE and AMDF on UKB disease classification and phenotype prediction under different synthetic-data mixing ratios.}
\label{Table5}
\resizebox{\textwidth}{!}{
\begin{tabular}{lcccccc|c}
\toprule
\multirow{2}{*}{\textbf{Model Variant}} 
& \multicolumn{2}{c}{\textbf{UKB-CAD}} 
& \multicolumn{2}{c}{\textbf{UKB-CM}} 
& \multicolumn{2}{c}{\textbf{UKB-HF}} 
& \multirow{2}{*}{\textbf{Overall $R^2$}} \\
\cmidrule(lr){2-3} \cmidrule(lr){4-5} \cmidrule(lr){6-7}
& \textbf{ACC$\uparrow$} 
& \textbf{AUC$\uparrow$} 
& \textbf{ACC$\uparrow$} 
& \textbf{AUC$\uparrow$} 
& \textbf{ACC$\uparrow$} 
& \textbf{AUC$\uparrow$} 
& \textbf{82 Phenotypes$\uparrow$} \\
\midrule

\multicolumn{8}{l}{\textit{Training with 100\% Mixed Synthetic Data}} \\
Baseline 
& \ablres{0.703}{0.021} & \ablres{0.765}{0.015} 
& \ablres{0.788}{0.028} & \ablres{0.814}{0.032} 
& \ablres{0.789}{0.019} & \ablres{0.862}{0.024} 
& $0.446$ \\
w/ PA-MAE 
& \ablres{0.708}{0.025} & \ablres{0.771}{0.018} 
& \ablres{0.792}{0.019} & \ablres{0.822}{0.028} 
& \ablres{0.794}{0.014} & \ablres{0.865}{0.017} 
& $0.451$ \\
w/ AMDF 
& \ablres{0.711}{0.028} & \ablres{0.778}{0.023} 
& \ablres{0.797}{0.021} & \ablres{0.828}{0.025} 
& \ablres{0.799}{0.023} & \ablres{0.871}{0.019} 
& $0.459$ \\
\rowcolor{gray!15}
Full Model 
& \ablbestres{0.716}{0.027} & \ablbestres{0.787}{0.011} 
& \ablbestres{0.806}{0.035} & \ablbestres{0.837}{0.024} 
& \ablbestres{0.808}{0.018} & \ablbestres{0.876}{0.026} 
& $\mathbf{0.470}$ \\
\midrule

\multicolumn{8}{l}{\textit{Training with 200\% Mixed Synthetic Data}} \\
Baseline 
& \ablres{0.709}{0.023} & \ablres{0.778}{0.031} 
& \ablres{0.799}{0.015} & \ablres{0.826}{0.034} 
& \ablres{0.801}{0.033} & \ablres{0.868}{0.025} 
& $0.451$ \\
w/ PA-MAE 
& \ablres{0.713}{0.016} & \ablres{0.785}{0.029} 
& \ablres{0.803}{0.011} & \ablres{0.831}{0.029} 
& \ablres{0.805}{0.025} & \ablres{0.872}{0.028} 
& $0.460$ \\
w/ AMDF 
& \ablres{0.717}{0.024} & \ablres{0.788}{0.023} 
& \ablres{0.808}{0.017} & \ablres{0.836}{0.036} 
& \ablres{0.809}{0.038} & \ablres{0.875}{0.030} 
& $0.466$ \\
\rowcolor{gray!15}
Full Model 
& \ablbestres{0.723}{0.019} & \ablbestres{0.796}{0.028} 
& \ablbestres{0.818}{0.013} & \ablbestres{0.845}{0.039} 
& \ablbestres{0.817}{0.030} & \ablbestres{0.886}{0.021} 
& $\mathbf{0.482}$ \\
\midrule

\multicolumn{8}{l}{\textit{Training with 300\% Mixed Synthetic Data}} \\
Baseline 
& \ablres{0.717}{0.018} & \ablres{0.785}{0.035} 
& \ablres{0.808}{0.023} & \ablres{0.836}{0.020} 
& \ablres{0.810}{0.025} & \ablres{0.871}{0.021} 
& $0.459$ \\
w/ PA-MAE 
& \ablres{0.721}{0.021} & \ablres{0.790}{0.028} 
& \ablres{0.814}{0.025} & \ablres{0.843}{0.021} 
& \ablres{0.816}{0.019} & \ablres{0.878}{0.024} 
& $0.468$ \\
w/ AMDF 
& \ablres{0.724}{0.015} & \ablres{0.796}{0.031} 
& \ablres{0.818}{0.021} & \ablres{0.845}{0.018} 
& \ablres{0.820}{0.023} & \ablres{0.881}{0.019} 
& $0.471$ \\
\rowcolor{gray!15}
Full Model 
& \ablbestres{0.730}{0.011} & \ablbestres{0.804}{0.033} 
& \ablbestres{0.826}{0.020} & \ablbestres{0.854}{0.014} 
& \ablbestres{0.826}{0.028} & \ablbestres{0.891}{0.016} 
& $\mathbf{0.499}$ \\
\bottomrule
\end{tabular}
}
\end{table*}

\subsection{Ablation Studies}
We conduct ablation studies on the UKB dataset to examine the contribution of the two key components in ECGFlowCMR. We consider three control variants: (1) a baseline model that encodes ECG signals with a standard MAE and generates latent videos using a DiT-based conditional flow-matching network initialized from Gaussian noise; (2) a PA-MAE variant that replaces the standard ECG encoder with PA-MAE while disabling AMDF; and (3) an AMDF variant that keeps the anatomy-motion disentangled flow design while using a standard MAE for ECG encoding.

\paragraph{{\bfseries Generation-level Ablation.}} 
We first evaluate whether each module improves the fidelity of generated cine CMR videos. As shown in Figure~\ref{Figure2}, the full model achieves the best performance across all generation metrics. Compared with the baseline, introducing PA-MAE improves the perceptual and distributional quality of generated videos, indicating that phase-aware ECG representations provide more informative cardiac-cycle conditioning. Introducing AMDF yields further gains, especially on FID and FVD, suggesting that explicitly disentangling static anatomy and dynamic motion helps produce more realistic and temporally coherent cine CMR videos. The full model consistently outperforms all variants, showing that PA-MAE and AMDF address complementary aspects of ECG-to-CMR generation.

\paragraph{{\bfseries Downstream-level Ablation.}} 
We further assess whether the generated videos preserve clinically useful information by training downstream evaluation models on synthetic data generated by each ablation variant. Table~\ref{Table5} reports disease classification and phenotype prediction results under different mixed synthetic data ratios. Across 100\%, 200\%, and 300\% mixed synthetic data settings, the full ECGFlowCMR model achieves the best overall performance on UKB-CAD, UKB-CM, UKB-HF, and the 82-phenotype prediction task. The PA-MAE and AMDF variants both improve over the baseline, confirming that phase-aware ECG encoding and anatomy-motion disentangled generation each contribute to clinically informative synthesis. The consistent advantage of the full model across synthetic data ratios indicates that combining PA-MAE with AMDF improves both visual fidelity and downstream utility, rather than only optimizing low-level video quality.

\subsection{Parameter Analysis}
\begin{figure}[t]
\centering
\includegraphics[width=0.99\linewidth]{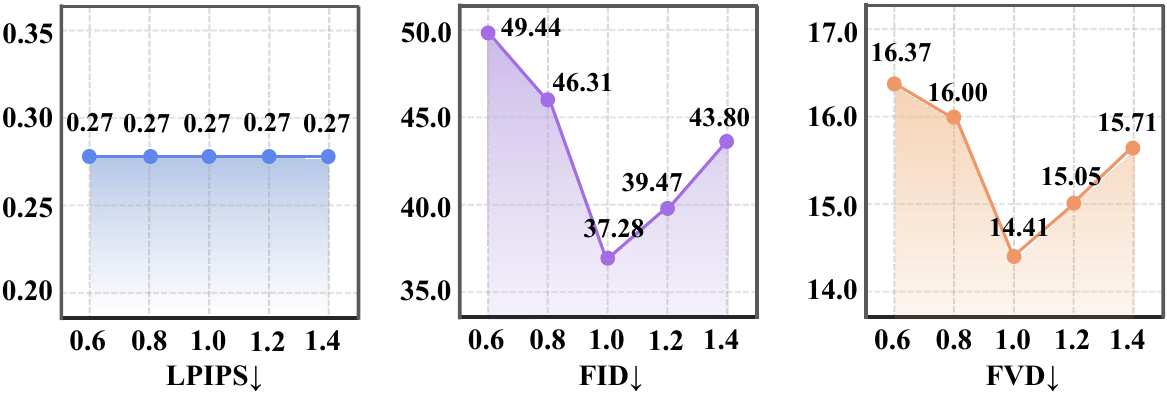}
\caption{Sensitivity analysis of the AMDF noise scale $\alpha$ on the UKB dataset.}
\label{Figure3}
\end{figure}

We further investigate the sensitivity of ECGFlowCMR to the noise scale hyperparameter $\alpha$ defined in Eq.~\ref{eq8}. As depicted in Figure~\ref{Figure3}, LPIPS remains stable across varying $\alpha$, whereas both FID and FVD exhibit a U-shaped response, attaining their optimal values at $\alpha = 1.0$. These results suggest that injecting an appropriate level of noise improves the distributional fidelity of the generated cine CMR sequences without compromising perceptual similarity.

\begin{figure*}[t]
\centering
\includegraphics[width=0.95\linewidth]{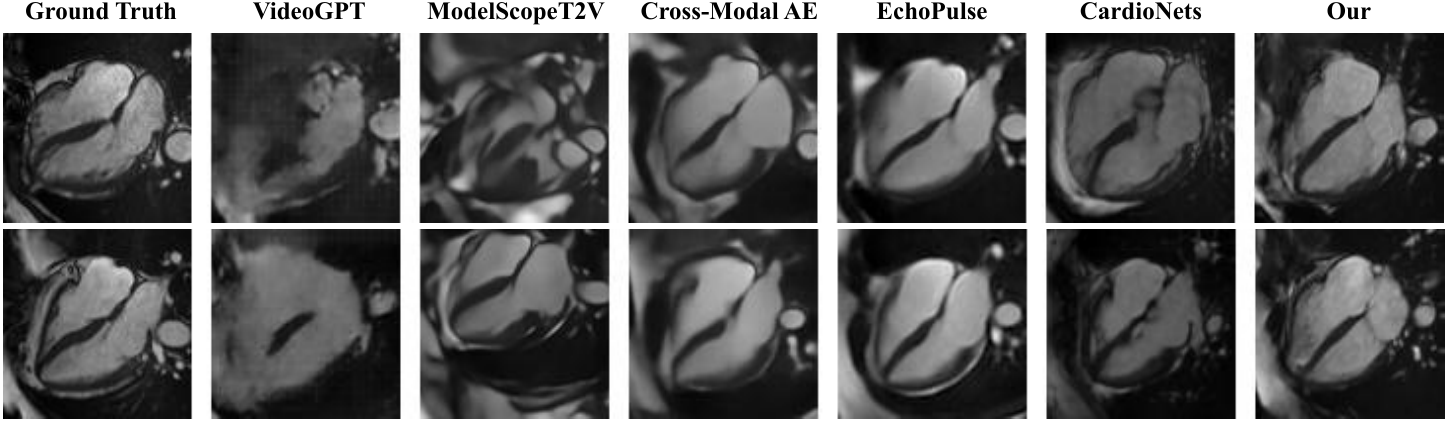}
\caption{Qualitative comparison of synthesized cine CMR frames on the UKB dataset. ECGFlowCMR produces sharper myocardial boundaries and more anatomically consistent ventricular morphology than competing generation methods.}
\label{Figure5}
\end{figure*}

\subsection{External Transfer Evaluation on ZJU-CM}
To evaluate generalization beyond the UKB distribution, we perform external transfer evaluation on ZJU-CM for binary and four-class cardiomyopathy classification. ECGFlowCMR is trained only on the UK Biobank (UKB) dataset. After training, the frozen model generates synthetic cine CMR sequences for the UKB training split, yielding augmented UKB pretraining sets with different synthetic-to-real mixing ratios. The MAE pretraining corpus is constructed exclusively from UKB data.

For each mixing ratio, we perform MAE-based self-supervised pretraining on the corresponding augmented UKB CMR set and transfer the pretrained encoder to ZJU-CM for downstream classification. On ZJU-CM, we use patient-level five-fold cross-validation across the full dataset. In each fold, patients are split into non-overlapping training and held-out evaluation folds, with all samples from the same patient assigned to the same fold. The model is fine-tuned on the ZJU-CM training folds and evaluated on held-out patients. This protocol is applied to both binary and four-class classification across all mixing ratios.

As shown in Table~\ref{Table4}, augmented UKB pretraining improves external classification performance up to an optimal synthetic-to-real ratio. For binary classification, the highest ACC (0.836) is achieved at 300\% mixing, and the highest AUC (0.848) is achieved at 400\%. For four-class classification, the best ACC (0.745) occurs at 400\%, while the highest AUC (0.853) occurs at 300\%. These results show that ECGFlowCMR-generated UKB augmentation improves the transferability of CMR representations to an external clinical cohort. The decline at larger mixing ratios suggests that excessive synthetic data may introduce redundancy or distributional bias during self-supervised pretraining, indicating the need to balance real and synthetic data for cross-domain transfer.

\begin{table}[t]
\small
\centering
\renewcommand\tabcolsep{5pt}
\renewcommand\arraystretch{1.0}
    \newcommand{\res}[2]{$#1_{\pm #2}$}
    
    \newcommand{\redres}[2]{\textcolor{red}{$#1_{\pm #2}$}}
    \newcommand{\bestres}[2]{\bm{$#1_{\pm #2}$}}
    
  \caption{External validations of ECGFlowCMR on the private ZJU-CM dataset under varying synthetic-data mixing ratios.}
  \begin{tabular}{lcc|cc}
    \toprule
    Classification & \multicolumn{2}{c}{\textbf{Binary Classification}} & \multicolumn{2}{c}{\textbf{Four-Class Classification}} \\
    \midrule
    Methods & \textbf{ACC$\uparrow$} & \textbf{AUC$\uparrow$} & \textbf{ACC$\uparrow$} & \textbf{AUC$\uparrow$} \\
    \midrule
    ViT~\cite{dosovitskiy2020image} & \res{0.736}{0.038} & \res{0.671}{0.061} & \res{0.531}{0.022} & \res{0.694}{0.055} \\
    MAE~\cite{he2022masked}(real) & \res{0.801}{0.054} & \res{0.798}{0.029} & \res{0.664}{0.047} & \res{0.784}{0.063} \\
    --mix 100\% & \res{0.804}{0.039} & \res{0.813}{0.041} & \res{0.687}{0.052} & \res{0.814}{0.028} \\
    --mix 200\% & \res{0.815}{0.047} & \res{0.826}{0.051} & \res{0.696}{0.035} & \res{0.834}{0.042} \\
    --mix 300\% & \bestres{0.836}{0.027} & \res{0.843}{0.029} & \res{0.717}{0.054} & \bestres{0.853}{0.026} \\
    --mix 400\% & \res{0.828}{0.043} & \bestres{0.848}{0.059} & \bestres{0.745}{0.045} & \res{0.846}{0.053} \\
    --mix 500\% & \res{0.818}{0.052} & \res{0.838}{0.037} & \res{0.733}{0.041} & \res{0.837}{0.055} \\
    \bottomrule
  \end{tabular}
  \label{Table4}
\end{table}

\subsection{Visualization Results}
Figure~\ref{Figure5} displays representative cine CMR frames generated by different models. ECGFlowCMR yields sharper myocardial boundaries and more anatomically consistent ventricular morphology. It also better preserves temporal coherence across cardiac phases, producing smoother motion and minimizing inter-frame artifacts. Additional video results are provided in our GitHub repository. 

\section{Limitations and Ethical Considerations}
ECGFlowCMR is trained and evaluated on four-chamber cine CMR data from the UK Biobank and one external clinical cohort. Consequently, its pretraining benefits may not generalize to other imaging views, scanner vendors, clinical sites, or rare pathophysiological conditions. The use of a population-level anatomical template and latent compression ensures stable anatomy-motion synthesis but may reduce adaptability to atypical morphologies, underscoring the need for multi-center and view-diverse training. All data usage must comply with IRB approval and informed consent, including strict de-identification and access control. The generated cine CMRs are intended solely for research applications, such as augmentation and representation learning, and are not approved for standalone clinical use. Clinical oversight remains essential to mitigate risks related to misuse, bias propagation, or privacy breaches.

\section{Conclusion}
In this paper, we propose ECGFlowCMR, a generative framework that synthesizes realistic cine CMR sequences from 12-lead ECGs by addressing two core challenges: cross-modal temporal mismatch and anatomical observability gap. Integrating the PA-MAE and the AMDF modules, ECGFlowCMR enables scalable pretraining and improves downstream performance in cardiac classification and phenotype prediction. Extensive experiments on the UK Biobank and a proprietary clinical dataset demonstrate its quantitative and qualitative superiority over prior methods, highlighting its potential for data-efficient cardiac imaging.

\section*{GenAI Disclosure}
During this research and manuscript preparation, we use Large Language Models (LLMs) for auxiliary tasks such as generating short scripts during coding and assisting with text translation and language polishing. All core ideas, research methodologies, and academic contributions are conceived and developed independently by the authors, with the role of LLMs limited to improving the fluency and readability of the presentation.

\section*{Acknowledgements}
This work is supported by the National Natural Science Foundation of China (62102008), CCF-Tencent Rhino-Bird Open Research Fund (CCF-Tencent RAGR20250108), CCF-Zhipu Large Model Innovation Fund (CCF-Zhipu202414), PKU-OPPO Fund (BO202301, BO202503), Research Project of Peking University in the State Key Laboratory of Vascular Homeostasis and Remodeling (2025-SKLVHR-YCTS-02), Beijing Municipal Science and Technology Commission (Z251100000725008), Prevention and Control of Emerging and Major Infectious Diseases-National Science and Technology Major Project (2025ZD01906000, 2025ZD01906004), Capital’s Funds for Health Improvement and Research (CFH2026-1-4092), Beijing Natural Science Foundation (QY26080). 

This research was conducted using the UK Biobank Resource under Application Number 90018. For the ZJU-CM dataset, the data collection protocols and the use of ECG, CMR, and clinical information were approved by the Institutional Review Board of the First Affiliated Hospital, Zhejiang University (FAHZU; Approval No. IIT20240507B).

The authors thank the cardiologists who participated in the expert evaluation: Guanyu Mu (The Second Hospital of Tianjin Medical University, Tianjin, China), Qinghao Zhao (Peking University People's Hospital, Beijing, China), Xingliang Wu (Tianjin Institute of Cardiology, Tianjin, China), Xinxin Di (The First Affiliated Hospital of USTC, Hefei, China), and Jing Zhao (The First Affiliated Hospital of Anhui Medical University, Hefei, China).


\bibliographystyle{ACM-Reference-Format}
\bibliography{sample-base}

@article{yan2021videogpt,
  title={Videogpt: Video generation using vq-vae and transformers},
  author={Yan, Wilson and Zhang, Yunzhi and Abbeel, Pieter and Srinivas, Aravind},
  journal={arXiv preprint arXiv:2104.10157},
  year={2021}
}

@article{wang2023modelscope,
  title={Modelscope text-to-video technical report},
  author={Wang, Jiuniu and Yuan, Hangjie and Chen, Dayou and Zhang, Yingya and Wang, Xiang and Zhang, Shiwei},
  journal={arXiv preprint arXiv:2308.06571},
  year={2023}
}

@inproceedings{liechopulse,
  title={ECHOPulse: ECG Controlled Echocardio-gram Video Generation},
  author={Li, Yiwei and Kim, Sekeun and Wu, Zihao and Jiang, Hanqi and Pan, Yi and Jin, Pengfei and Song, Sifan and Shi, Yucheng and Yu, Xiaowei and Yang, Tianze and others},
  booktitle={ICLR},
  year={2025}
}

@article{dosovitskiy2020image,
  title={An image is worth 16x16 words: Transformers for image recognition at scale},
  author={Dosovitskiy, Alexey},
  journal={arXiv preprint arXiv:2010.11929},
  year={2020}
}

@inproceedings{he2022masked,
  title={Masked autoencoders are scalable vision learners},
  author={He, Kaiming and Chen, Xinlei and Xie, Saining and Li, Yanghao and Doll{\'a}r, Piotr and Girshick, Ross},
  booktitle={CVPR},
  pages={16000--16009},
  year={2022}
}

@article{bai2020population,
  title={A population-based phenome-wide association study of cardiac and aortic structure and function},
  author={Bai, Wenjia and Suzuki, Hideaki and Huang, Jian and Francis, Catherine and Wang, Shuo and Tarroni, Giacomo and Guitton, Florian and Aung, Nay and Fung, Kenneth and Petersen, Steffen E and others},
  journal={Nature medicine},
  volume={26},
  number={10},
  pages={1654--1662},
  year={2020},
  publisher={Nature Publishing Group US New York}
}

@article{wang2024screening,
  title={Screening and diagnosis of cardiovascular disease using artificial intelligence-enabled cardiac magnetic resonance imaging},
  author={Wang, Yan-Ran and Yang, Kai and Wen, Yi and Wang, Pengcheng and Hu, Yuepeng and Lai, Yongfan and Wang, Yufeng and Zhao, Kankan and Tang, Siyi and Zhang, Angela and others},
  journal={Nature Medicine},
  volume={30},
  number={5},
  pages={1471--1480},
  year={2024},
  publisher={Nature Publishing Group US New York}
}

@article{bluethgen2025vision,
  title={A vision--language foundation model for the generation of realistic chest x-ray images},
  author={Bluethgen, Christian and Chambon, Pierre and Delbrouck, Jean-Benoit and Van Der Sluijs, Rogier and Po{\l}acin, Ma{\l}gorzata and Zambrano Chaves, Juan Manuel and Abraham, Tanishq Mathew and Purohit, Shivanshu and Langlotz, Curtis P and Chaudhari, Akshay S},
  journal={Nature Biomedical Engineering},
  volume={9},
  number={4},
  pages={494--506},
  year={2025},
  publisher={Nature Publishing Group UK London}
}

@article{tudosiu2024realistic,
  title={Realistic morphology-preserving generative modelling of the brain},
  author={Tudosiu, Petru-Daniel and Pinaya, Walter HL and Ferreira Da Costa, Pedro and Dafflon, Jessica and Patel, Ashay and Borges, Pedro and Fernandez, Virginia and Graham, Mark S and Gray, Robert J and Nachev, Parashkev and others},
  journal={Nature Machine Intelligence},
  volume={6},
  number={7},
  pages={811--819},
  year={2024},
  publisher={Nature Publishing Group UK London}
}

@article{zakeri2023dragnet,
  title={DragNet: Learning-based deformable registration for realistic cardiac MR sequence generation from a single frame},
  author={Zakeri, Arezoo and Hokmabadi, Alireza and Bi, Ning and Wijesinghe, Isuru and Nix, Michael G and Petersen, Steffen E and Frangi, Alejandro F and Taylor, Zeike A and Gooya, Ali},
  journal={Medical Image Analysis},
  volume={83},
  pages={102678},
  year={2023},
  publisher={Elsevier}
}

@inproceedings{li2025phenotype,
  title={Phenotype-Guided Generative Model for High-Fidelity Cardiac MRI Synthesis: Advancing Pretraining and Clinical Applications},
  author={Li, Ziyu and Hu, Yujian and Ding, Zhengyao and Mao, Yiheng and Li, Haitao and Yi, Fan and Zhang, Hongkun and Huang, Zhengxing},
  booktitle={MICCAI},
  pages={484--494},
  year={2025},
  organization={Springer}
}

@inproceedings{zhang2018unreasonable,
  title={The unreasonable effectiveness of deep features as a perceptual metric},
  author={Zhang, Richard and Isola, Phillip and Efros, Alexei A and Shechtman, Eli and Wang, Oliver},
  booktitle={CVPR},
  pages={586--595},
  year={2018}
}

@inproceedings{heusel2017gans,
  title={Gans trained by a two time-scale update rule converge to a local nash equilibrium},
  author={Heusel, Martin and Ramsauer, Hubert and Unterthiner, Thomas and Nessler, Bernhard and Hochreiter, Sepp},
  booktitle={NeurIPS},
  volume={30},
  year={2017}
}

@article{unterthiner2019fvd,
  title={FVD: A new metric for video generation},
  author={Unterthiner, Thomas and Van Steenkiste, Sjoerd and Kurach, Karol and Marinier, Rapha{\"e}l and Michalski, Marcin and Gelly, Sylvain},
  year={2019}
}

@inproceedings{abbaspourazadlarge,
  title={Large-scale Training of Foundation Models for Wearable Biosignals},
  author={Abbaspourazad, Salar and Elachqar, Oussama and Miller, Andrew and Emrani, Saba and Nallasamy, Udhyakumar and Shapiro, Ian},
  booktitle={ICLR},
  year={2024}
}

@article{radhakrishnan2023cross,
  title={Cross-modal autoencoder framework learns holistic representations of cardiovascular state},
  author={Radhakrishnan, Adityanarayanan and Friedman, Sam F and Khurshid, Shaan and Ng, Kenney and Batra, Puneet and Lubitz, Steven A and Philippakis, Anthony A and Uhler, Caroline},
  journal={Nature Communications},
  volume={14},
  number={1},
  pages={2436},
  year={2023},
  publisher={Nature Publishing Group UK London}
}

@article{bycroft2018uk,
  title={The UK Biobank resource with deep phenotyping and genomic data},
  author={Bycroft, Clare and Freeman, Colin and Petkova, Desislava and Band, Gavin and Elliott, Lloyd T and Sharp, Kevin and Motyer, Allan and Vukcevic, Damjan and Delaneau, Olivier and O’Connell, Jared and others},
  journal={Nature},
  volume={562},
  number={7726},
  pages={203--209},
  year={2018},
  publisher={Nature Publishing Group UK London}
}

@inproceedings{prakash2025evaluating,
  title={Evaluating and Improving the Effectiveness of Synthetic Chest X-Rays for Medical Image Analysis},
  author={Prakash, Eva and Valanarasu, Jeya Maria Jose and Chen, Zhihong and Reis, Eduardo Pontes and Johnston, Andrew and Pareek, Anuj and Bluethgen, Christian and Gatidis, Sergios and Olsen, Cameron and Chaudhari, Akshay S and others},
  booktitle={ICCV},
  pages={4413--4421},
  year={2025}
}

@inproceedings{peng2023generating,
  title={Generating realistic brain mris via a conditional diffusion probabilistic model},
  author={Peng, Wei and Adeli, Ehsan and Bosschieter, Tomas and Park, Sang Hyun and Zhao, Qingyu and Pohl, Kilian M},
  booktitle={MICCAI},
  pages={14--24},
  year={2023},
  organization={Springer}
}

@inproceedings{chen2024ultrasound,
  title={Ultrasound image-to-video synthesis via latent dynamic diffusion models},
  author={Chen, Tingxiu and Shi, Yilei and Zheng, Zixuan and Yan, Bingcong and Hu, Jingliang and Zhu, Xiao Xiang and Mou, Lichao},
  booktitle={MICCAI},
  pages={764--774},
  year={2024},
  organization={Springer}
}

@inproceedings{liuflow,
  title={Flow Straight and Fast: Learning to Generate and Transfer Data with Rectified Flow},
  author={Liu, Xingchao and Gong, Chengyue and others},
  booktitle={ICLR},
  year={2023}
}

@inproceedings{peebles2023scalable,
  title={Scalable diffusion models with transformers},
  author={Peebles, William and Xie, Saining},
  booktitle={ICCV},
  pages={4195--4205},
  year={2023}
}

@article{moor2023foundation,
  title={Foundation models for generalist medical artificial intelligence},
  author={Moor, Michael and Banerjee, Oishi and Abad, Zahra Shakeri Hossein and Krumholz, Harlan M and Leskovec, Jure and Topol, Eric J and Rajpurkar, Pranav},
  journal={Nature},
  volume={616},
  number={7956},
  pages={259--265},
  year={2023},
  publisher={Nature Publishing Group UK London}
}

@article{vondrick2016generating,
  title={Generating videos with scene dynamics},
  author={Vondrick, Carl and Pirsiavash, Hamed and Torralba, Antonio},
  journal={Advances in neural information processing systems},
  volume={29},
  year={2016}
}

@inproceedings{saito2017temporal,
  title={Temporal generative adversarial nets with singular value clipping},
  author={Saito, Masaki and Matsumoto, Eiichi and Saito, Shunta},
  booktitle={Proceedings of the IEEE international conference on computer vision},
  pages={2830--2839},
  year={2017}
}

@inproceedings{tulyakov2018mocogan,
  title={Mocogan: Decomposing motion and content for video generation},
  author={Tulyakov, Sergey and Liu, Ming-Yu and Yang, Xiaodong and Kautz, Jan},
  booktitle={Proceedings of the IEEE conference on computer vision and pattern recognition},
  pages={1526--1535},
  year={2018}
}

@article{babaeizadeh2017stochastic,
  title={Stochastic variational video prediction},
  author={Babaeizadeh, Mohammad and Finn, Chelsea and Erhan, Dumitru and Campbell, Roy H and Levine, Sergey},
  journal={arXiv preprint arXiv:1710.11252},
  year={2017}
}

@inproceedings{denton2018stochastic,
  title={Stochastic video generation with a learned prior},
  author={Denton, Emily and Fergus, Rob},
  booktitle={International conference on machine learning},
  pages={1174--1183},
  year={2018},
  organization={PMLR}
}

@article{van2017neural,
  title={Neural discrete representation learning},
  author={Van Den Oord, Aaron and Vinyals, Oriol and others},
  journal={Advances in neural information processing systems},
  volume={30},
  year={2017}
}

@article{villegas2022phenaki,
  title={Phenaki: Variable length video generation from open domain textual description},
  author={Villegas, Ruben and Babaeizadeh, Mohammad and Kindermans, Pieter-Jan and Moraldo, Hernan and Zhang, Han and Saffar, Mohammad Taghi and Castro, Santiago and Kunze, Julius and Erhan, Dumitru},
  journal={arXiv preprint arXiv:2210.02399},
  year={2022}
}

@article{ho2022video,
  title={Video diffusion models},
  author={Ho, Jonathan and Salimans, Tim and Gritsenko, Alexey and Chan, William and Norouzi, Mohammad and Fleet, David J},
  journal={Advances in neural information processing systems},
  volume={35},
  pages={8633--8646},
  year={2022}
}

@article{he2022latent,
  title={Latent video diffusion models for high-fidelity long video generation},
  author={He, Yingqing and Yang, Tianyu and Zhang, Yong and Shan, Ying and Chen, Qifeng},
  journal={arXiv preprint arXiv:2211.13221},
  year={2022}
}

@inproceedings{blattmann2023align,
  title={Align your latents: High-resolution video synthesis with latent diffusion models},
  author={Blattmann, Andreas and Rombach, Robin and Ling, Huan and Dockhorn, Tim and Kim, Seung Wook and Fidler, Sanja and Kreis, Karsten},
  booktitle={Proceedings of the IEEE/CVF conference on computer vision and pattern recognition},
  pages={22563--22575},
  year={2023}
}

@article{singer2022make,
  title={Make-a-video: Text-to-video generation without text-video data},
  author={Singer, Uriel and Polyak, Adam and Hayes, Thomas and Yin, Xi and An, Jie and Zhang, Songyang and Hu, Qiyuan and Yang, Harry and Ashual, Oron and Gafni, Oran and others},
  journal={arXiv preprint arXiv:2209.14792},
  year={2022}
}

@inproceedings{vukadinovic2023gancmri,
  title={GANcMRI: Cardiac magnetic resonance video generation and physiologic guidance using latent space prompting},
  author={Vukadinovic, Milos and Kwan, Alan C and Li, Debiao and Ouyang, David},
  booktitle={Machine Learning for Health (ML4H)},
  pages={594--606},
  year={2023},
  organization={PMLR}
}

@inproceedings{liu2024texdc,
  title={Texdc: Text-driven disease-aware 4d cardiac cine mri images generation},
  author={Liu, Cong and Yuan, Xiaohan and Yu, ZhiPeng and Wang, Yangang},
  booktitle={Proceedings of the Asian Conference on Computer Vision},
  pages={3005--3021},
  year={2024}
}

@article{lei2025patch,
  title={A patch-based latent video diffusion model for cardiac cine (Cine-LDiff)},
  author={Lei, Xuan and Schniter, Philip and Ahmad, Rizwan},
  journal={Journal of Cardiovascular Magnetic Resonance},
  volume={27},
  year={2025},
  publisher={Elsevier}
}

@inproceedings{liu2024tagged,
  title={Tagged-to-Cine MRI Sequence Synthesis via Light Spatial-Temporal Transformer},
  author={Liu, Xiaofeng and Xing, Fangxu and Bian, Zhangxing and Arias-Vergara, Tomas and P{\'e}rez-Toro, Paula Andrea and Maier, Andreas and Stone, Maureen and Zhuo, Jiachen and Prince, Jerry L and Woo, Jonghye},
  booktitle={International Conference on Medical Image Computing and Computer-Assisted Intervention},
  pages={701--711},
  year={2024},
  organization={Springer}
}

@inproceedings{deb2025unsupervised,
  title={Unsupervised Cardiac Video Translation Via Motion Feature Guided Diffusion Model},
  author={Deb, Swakshar and Wu, Nian and Epstein, Frederick H and Zhang, Miaomiao},
  booktitle={International Conference on Medical Image Computing and Computer-Assisted Intervention},
  pages={648--658},
  year={2025},
  organization={Springer}
}

@article{ding2026generating,
  title={Generating Cardiac Magnetic Resonance Images from Electrocardiograms—A Multicenter Study},
  author={Ding, Zhengyao and Li, Ziyu and Hu, Yujian and Xu, Youyao and Zhao, Chengchen and Mao, Yiheng and Li, Haitao and Li, Zhikang and Li, Qian and Wang, Jing and others},
  journal={NEJM AI},
  volume={3},
  number={4},
  pages={AIoa2500549},
  year={2026},
  publisher={Massachusetts Medical Society}
}

@article{nie2025anyppg,
  title={Anyppg: An ecg-guided ppg foundation model trained on over 100,000 hours of recordings for holistic health profiling},
  author={Nie, Guangkun and Tang, Gongzheng and Xiao, Yujie and Li, Jun and Huang, Shun and Zhang, Deyun and Zhao, Qinghao and Hong, Shenda},
  journal={arXiv preprint arXiv:2511.01747},
  year={2025}
}

@article{zhang2026ecgomics,
  title={ECGomics: An open platform for AI-ECG digital biomarker discovery},
  author={Zhang, Deyun and Li, Jun and Geng, Shijia and Wang, Yue and Chen, Shijie and Fan, Sumei and Zhao, Qinghao and Hong, Shenda},
  journal={Health Data Science},
  volume={6},
  pages={0427},
  year={2026},
  publisher={AAAS}
}

@article{jin2026ecg,
  title={ECG-R1: Protocol-Guided and Modality-Agnostic MLLM for Reliable ECG Interpretation},
  author={Jin, Jiarui and Wang, Haoyu and Wu, Xingliang and Fang, Xiaocheng and Lan, Xiang and Wang, Zihan and Zhang, Deyun and Liu, Bo and Zhang, Yingying and Wu, Xian and others},
  journal={arXiv preprint arXiv:2602.04279},
  year={2026}
}

@article{jin2025self,
  title={Self-Alignment Learning to Improve Myocardial Infarction Detection from Single-Lead ECG},
  author={Jin, Jiarui and Fang, Xiaocheng and Wang, Haoyu and Li, Jun and Liu, Che and Xie, Donglin and Li, Hongyan and Hong, Shenda},
  journal={arXiv preprint arXiv:2509.19397},
  year={2025}
}

@article{fang2025ppgflowecg,
  title={PPGFlowECG: Latent Rectified Flow with Cross-Modal Encoding for PPG-Guided ECG Generation and Cardiovascular Disease Detection},
  author={Fang, Xiaocheng and Jin, Jiarui and Wang, Haoyu and Liu, Che and Cai, Jieyi and Xiao, Yujie and Nie, Guangkun and Liu, Bo and Huang, Shun and Li, Hongyan and others},
  journal={arXiv preprint arXiv:2509.19774},
  year={2025}
}

@inproceedings{wang2026se,
  title={SE-Diff: Simulator and Experience Enhanced Diffusion Model for Comprehensive ECG Generation},
  author={Wang, Xiaoda and Han, Kaiqiao and Xu, Yuhao and Luo, Xiao and Sun, Yizhou and Wang, Wei and Yang, Carl},
  booktitle={The Fourteenth International Conference on Learning Representations},
  year={2026}
}

@article{bao2026position,
  title={Position: General Alignment Has Hit a Ceiling; Edge Alignment Must Be Taken Seriously},
  author={Bao, Han and Huang, Yue and Wang, Xiaoda and Zhang, Zheyuan and Zhou, Yujun and Yang, Carl and Zhang, Xiangliang and Ye, Yanfang},
  journal={arXiv e-prints},
  pages={arXiv--2602},
  year={2026}
}

\appendix
\section{Algorithm of ECGFlowCMR}

\begin{algorithm}[h]
\caption{Phase-Aware Masked Autoencoder}
\label{alg:pamae}
\begin{algorithmic}[1]
\State \textbf{Input:} ECG dataset $\mathcal{D}_{\mathrm{ECG}}$, masking ratio $\rho$.
\State \textbf{Ensure:} Optimized PA-MAE parameters $\theta_{\mathrm{PA}}$.
\While{not converged}
    \State Sample a mini-batch $x_{\mathrm{ECG}}\sim\mathcal{D}_{\mathrm{ECG}}$.
    \State Extract ECG features: $F\leftarrow E_{\mathrm{ECG}}(x_{\mathrm{ECG}})$.
    \State Apply temporal masking: $F_{\mathrm{masked}}\leftarrow F\odot m$, where $m$ is sampled with ratio $\rho$.
    \State Reconstruct ECG: $\widehat{x}_{\mathrm{ECG}}\leftarrow D_{\mathrm{ECG}}(F_{\mathrm{masked}})$.
    \State Predict cardiac phase: $\widehat{\Phi}\leftarrow P_{\mathrm{ECG}}(F)=[\sin(\phi),\cos(\phi)]$.
    \State Compute phase label $\Phi_{\mathrm{gt}}$ by Lead-II R-peak detection and R--R interpolation.
    \State $\mathcal{L}_{\mathrm{rec}}\leftarrow \frac{1}{CT}\|\widehat{x}_{\mathrm{ECG}}-x_{\mathrm{ECG}}\|_2^2$.
    \State $\mathcal{L}_{\mathrm{phase}}\leftarrow \frac{1}{T'}\|\widehat{\Phi}-\Phi_{\mathrm{gt}}\|_2^2$.
    \State $\mathcal{L}_{\mathrm{PAMAE}}\leftarrow \mathcal{L}_{\mathrm{rec}}+\mathcal{L}_{\mathrm{phase}}$.
    \State Update $\theta_{\mathrm{PA}}$ using $\nabla\mathcal{L}_{\mathrm{PAMAE}}$.
\EndWhile
\end{algorithmic}
\end{algorithm}

\begin{algorithm}[h]
\caption{Anatomy-Motion Disentangled Flow}
\label{alg:amdf}
\begin{algorithmic}[1]
\State \textbf{Input:} Paired dataset $\mathcal{D}$ of $(x_{\mathrm{ECG}},x_{\mathrm{CMR}})$, frozen $E_{\mathrm{ECG}}$ and $P_{\mathrm{ECG}}$, trained CMR encoder $E_{\mathrm{CMR}}$, noise scale $\alpha$, flow model $v_\theta$.
\State \textbf{Ensure:} Optimized AMDF parameters $\theta$.
\State Encode all training CMR videos $z_i\leftarrow E_{\mathrm{CMR}}(x_{\mathrm{CMR}}^{(i)})$ and compute $z_{\mathrm{template}}\leftarrow \frac{1}{NT}\sum_{i=1}^{N}\sum_{t=1}^{T}z_{i,t}$.
\While{not converged}
    \State Sample a mini-batch $(x_{\mathrm{ECG}},x_{\mathrm{CMR}})\sim\mathcal{D}$.
    \State $F\leftarrow E_{\mathrm{ECG}}(x_{\mathrm{ECG}})$, $\widehat{\Phi}\leftarrow P_{\mathrm{ECG}}(F)$.
    \State $c\leftarrow \mathrm{ROIAlign}(\mathrm{CycleDetect}(F,\widehat{\Phi}),50)$.
    \State $z_1\leftarrow E_{\mathrm{CMR}}(x_{\mathrm{CMR}})$.
    \State $z_0\leftarrow z_{\mathrm{template}}+\alpha\epsilon$, where $\epsilon\sim\mathcal{N}(\mathbf{0},\mathbf{I})$.
    \State $z_t\leftarrow (1-t)z_0+t z_1$, where $t\sim\mathcal{U}(0,1)$.
    \State $\mathcal{L}_{\mathrm{AMDF}}\leftarrow \|v_\theta(z_t,t,c)-(z_1-z_0)\|_2^2$.
    \State Update $\theta$ using $\nabla\mathcal{L}_{\mathrm{AMDF}}$.
\EndWhile
\end{algorithmic}
\end{algorithm}

\section{Additional Dataset Details}
\subsection{Data Splits and Evaluation Protocols}
\paragraph{{\bfseries ECG-to-CMR generation.}}
We trained ECGFlowCMR to synthesize cine CMR sequences from ECG signals using patient-disjoint training, validation, and test splits containing 29,490, 4,212, and 8,427 sequences, respectively. These splits were used for model training, validation, and evaluation while maintaining patient-level separation across partitions. After training, we froze ECGFlowCMR and used it as a data generator for downstream augmentation. We applied the frozen model to the training split with different random seeds to generate synthetic data at multiple expansion levels.

\paragraph{{\bfseries Cardiac disease classification.}}
For downstream disease classification, we constructed three balanced UKB cohorts with equal numbers of positive and negative samples: Coronary Artery Disease (UKB-CAD, n=5,464), Cardiomyopathy (UKB-CM, n=196), and Heart Failure (UKB-HF, n=578). Positive samples were defined by UKB diagnosis labels, and negative samples were selected from patients without the corresponding target disease. On the ECG-to-CMR training split, we first performed MAE-based self-supervised pretraining on CMR sequences. In the augmented setting, synthetic cine CMR sequences generated by the frozen ECGFlowCMR model were added to the CMR corpus used for MAE-based self-supervised pretraining. Within each disease cohort, five-fold cross-validation was used for fine-tuning and hyperparameter selection, while reserved samples were held out for final evaluation. ECGFlowCMR remained frozen throughout downstream training, and disease labels were used only for the corresponding classification task.

\paragraph{{\bfseries Cardiac phenotype prediction.}}
For downstream phenotype prediction, we retained ECG-to-CMR samples with complete phenotype annotations, yielding 19,487 training, 2,758 validation, and 5,553 test sequences across 82 cardiac phenotypes. Following the classification setting, we used the ECG-to-CMR training split for MAE-based self-supervised pretraining. For augmentation, synthetic cine CMR sequences generated by the frozen ECGFlowCMR model with different random seeds were added to the MAE pretraining corpus to construct expanded pretraining sets. Since the phenotype-labeled dataset provides sufficient supervised samples, models were fine-tuned directly on the training set, selected on the validation set, and evaluated on the independent test set.

\subsection{Task-Specific Cohorts and Split Integrity}
The disease classification cohorts and the phenotype-prediction subset were constructed from the same UKB resource using task-specific inclusion criteria. For disease classification, each cohort was defined according to the presence or absence of a target diagnosis, resulting in balanced cohorts for UKB-CAD, UKB-CM, and UKB-HF. For phenotype prediction, the subset was filtered to include sequences with complete annotations for all 82 cardiac phenotypes. Consequently, a patient may be included in both downstream task families when satisfying the corresponding criteria.

Each downstream task was trained and evaluated independently with its own label space, model selection protocol, and held-out evaluation split. All train, validation, and test partitions were constructed at the patient level, ensuring that samples from the same patient were assigned to the same partition within each task. The ECGFlowCMR model was trained separately and kept fixed during downstream augmentation, while disease labels and phenotype labels were used only within their corresponding downstream tasks. This protocol preserves split-level data isolation and prevents supervision from one downstream task from influencing another.

\begin{table}[t]
\scriptsize
\centering
\setlength{\tabcolsep}{3pt}
\renewcommand{\arraystretch}{1.0}
\caption{External transfer evaluation on the private ZJU-CM dataset using MAE models pretrained on UKB data augmented with different synthetic-data generators.}
\label{tab:zju_external_generator_comparison}
\resizebox{\columnwidth}{!}{%
\begin{tabular}{lcc|cc}
\toprule
\multirow{2}{*}{\textbf{Method}}
& \multicolumn{2}{c}{\textbf{Binary}} 
& \multicolumn{2}{c}{\textbf{Four-Class}} \\
\cmidrule(lr){2-3} \cmidrule(lr){4-5}
& \textbf{ACC$\uparrow$}
& \textbf{AUC$\uparrow$}
& \textbf{ACC$\uparrow$}
& \textbf{AUC$\uparrow$} \\
\midrule
ViT~\cite{dosovitskiy2020image}
& $0.736{\pm}0.038$ & $0.671{\pm}0.061$
& $0.531{\pm}0.022$ & $0.694{\pm}0.055$ \\
MAE~\cite{he2022masked} (real)
& $0.801{\pm}0.054$ & $0.798{\pm}0.029$
& $0.664{\pm}0.047$ & $0.784{\pm}0.063$ \\
\midrule
\multicolumn{5}{l}{\textit{100\% Mixed Synthetic Data}} \\
VideoGPT~\cite{yan2021videogpt}
& $0.792{\pm}0.041$ & $0.795{\pm}0.033$
& $0.659{\pm}0.046$ & $0.783{\pm}0.039$ \\
ModelScopeT2V~\cite{wang2023modelscope}
& $0.794{\pm}0.045$ & $0.797{\pm}0.037$
& $0.663{\pm}0.041$ & $0.785{\pm}0.052$ \\
EchoPulse~\cite{liechopulse}
& $0.799{\pm}0.028$ & $0.805{\pm}0.032$
& $0.675{\pm}0.043$ & $0.801{\pm}0.044$ \\
CardioNets~\cite{ding2026generating}
& $0.797{\pm}0.032$ & $0.801{\pm}0.029$
& $0.671{\pm}0.036$ & $0.793{\pm}0.035$ \\
\rowcolor{gray!15}
ECGFlowCMR
& $\mathbf{0.804{\pm}0.039}$ & $\mathbf{0.813{\pm}0.041}$
& $\mathbf{0.687{\pm}0.052}$ & $\mathbf{0.814{\pm}0.028}$ \\
\midrule
\multicolumn{5}{l}{\textit{200\% Mixed Synthetic Data}} \\
VideoGPT~\cite{yan2021videogpt}
& $0.801{\pm}0.033$ & $0.804{\pm}0.022$
& $0.668{\pm}0.027$ & $0.789{\pm}0.041$ \\
ModelScopeT2V~\cite{wang2023modelscope}
& $0.803{\pm}0.038$ & $0.808{\pm}0.031$
& $0.673{\pm}0.039$ & $0.804{\pm}0.025$ \\
EchoPulse~\cite{liechopulse}
& $0.807{\pm}0.046$ & $0.815{\pm}0.044$
& $0.688{\pm}0.033$ & $0.817{\pm}0.039$ \\
CardioNets~\cite{ding2026generating}
& $0.806{\pm}0.041$ & $0.811{\pm}0.035$
& $0.681{\pm}0.043$ & $0.813{\pm}0.034$ \\
\rowcolor{gray!15}
ECGFlowCMR
& $\mathbf{0.815{\pm}0.047}$ & $\mathbf{0.826{\pm}0.051}$
& $\mathbf{0.696{\pm}0.035}$ & $\mathbf{0.834{\pm}0.042}$ \\
\midrule
\multicolumn{5}{l}{\textit{300\% Mixed Synthetic Data}} \\
VideoGPT~\cite{yan2021videogpt}
& $0.810{\pm}0.041$ & $0.814{\pm}0.034$
& $0.675{\pm}0.039$ & $0.799{\pm}0.031$ \\
ModelScopeT2V~\cite{wang2023modelscope}
& $0.816{\pm}0.037$ & $0.819{\pm}0.028$
& $0.682{\pm}0.030$ & $0.818{\pm}0.029$ \\
EchoPulse~\cite{liechopulse}
& $0.822{\pm}0.043$ & $0.826{\pm}0.033$
& $0.698{\pm}0.025$ & $0.829{\pm}0.040$ \\
CardioNets~\cite{ding2026generating}
& $0.820{\pm}0.031$ & $0.823{\pm}0.039$
& $0.699{\pm}0.036$ & $0.831{\pm}0.033$ \\
\rowcolor{gray!15}
ECGFlowCMR
& $\mathbf{0.836{\pm}0.027}$ & $\mathbf{0.843{\pm}0.029}$
& $\mathbf{0.717{\pm}0.054}$ & $\mathbf{0.853{\pm}0.026}$ \\
\bottomrule
\end{tabular}%
}
\end{table}

\subsection{ZJU-CM External Validation Dataset}

The ZJU-CM dataset was collected from a clinical cardiac MRI cohort acquired on a 3.0T Philips Ingenia Elition X system with multi-channel phased-array coils. Cine CMR imaging was performed using a breath-hold, ECG-gated balanced steady-state free precession sequence, referred to as Philips B-TFE and clinically equivalent to bSSFP/Cine. The dataset contains four-chamber long-axis views (A4CH), providing cardiac motion sequences for external evaluation beyond the UK Biobank cohort.

The main acquisition parameters were as follows: repetition time $\mathrm{TR}=3.04$ ms, echo time $\mathrm{TE}=1.52$ ms, flip angle $45^\circ$, acquisition matrix $188\times175$, field of view $300$ mm, reconstructed pixel spacing approximately $1.04$ mm, slice thickness $8$ mm, no inter-slice gap, and SENSE parallel acquisition factor $2$. Each sequence was acquired within a single breath-hold of approximately $6.83$ s. ECG gating was performed with an R-wave trigger delay of $408$ ms. The heart rate during acquisition was approximately $82$ bpm, corresponding to an R-R interval of $675$--$776$ ms, with $8$ cardiac phases per cycle.

Image reconstruction used RIESZ k-space filtering followed by pixel-value normalization to 16-bit grayscale images with an effective 12-bit dynamic range. To align the preprocessing pipeline with the UK Biobank dataset, we applied a cardiac segmentation model~\cite{bai2020population} to localize the cardiac region and resized the resulting volumes to $50\times96\times96$. Downstream labels were derived from the institutional electronic medical record system.

\section{Additional Experimental Results}

\subsection{Comparison with Generative Baselines on ZJU-CM}
Table~\ref{tab:zju_external_generator_comparison} reports an external transfer comparison on ZJU-CM dataset. ViT serves as the supervised baseline, and MAE pretrained on real UKB CMR serves as the self-supervised baseline. This comparison evaluates whether self-supervised CMR pretraining improves transfer over supervised training and whether adding generated cine CMR sequences to the MAE pretraining corpus further improves external performance.

We further compare ECGFlowCMR with representative video and cardiac generation baselines under the same synthetic-to-real mixing ratios. For each generator, synthetic cine CMR sequences are added to the UKB pretraining set, followed by MAE pretraining on the augmented UKB data. The pretrained encoder is then transferred to ZJU-CM and fine-tuned using patient-level five-fold cross-validation. Across the evaluated mixing ratios and classification settings, ECGFlowCMR achieves the best overall performance, suggesting that its generated cine CMR sequences provide more transferable representations for external clinical evaluation.

\begin{table}[t]
\small
\centering
\setlength{\tabcolsep}{3pt}
\renewcommand{\arraystretch}{1.05}

\newcommand{\expertci}[3]{%
\begin{tabular}[c]{@{}c@{}}
$#1$\\
{\scriptsize [$#2$, $#3$]}
\end{tabular}}

\caption{Expert-level comparison between CardioNets and ECGFlowCMR. Avg. denotes the average score across five experts.}
\label{tab:expert_comparison}
\resizebox{\columnwidth}{!}{%
\begin{tabular}{lcccccc}
\toprule
\multirow{2}{*}{\textbf{Method}}
& \multirow{2}{*}{\textbf{Avg.}}
& \multicolumn{5}{c}{\textbf{Experts}} \\
\cmidrule(lr){3-7}
& 
& \textbf{Expert 1}
& \textbf{Expert 2}
& \textbf{Expert 3}
& \textbf{Expert 4}
& \textbf{Expert 5} \\
\midrule
CardioNets
& $0.588$
& \expertci{0.56}{0.45}{0.66}
& \expertci{0.58}{0.48}{0.68}
& \expertci{0.71}{0.62}{0.79}
& \expertci{0.57}{0.47}{0.67}
& \expertci{0.52}{0.42}{0.62} \\
ECGFlowCMR
& $\mathbf{0.514}$
& \expertci{0.52}{0.42}{0.62}
& \expertci{0.53}{0.43}{0.63}
& \expertci{0.60}{0.50}{0.69}
& \expertci{0.52}{0.42}{0.62}
& \expertci{0.40}{0.31}{0.50} \\
\bottomrule
\end{tabular}%
}
\end{table}

\subsection{Expert Realism Assessment of Synthetic Cine CMR Videos}

We conducted a human expert study to evaluate the perceptual realism of synthetic cine CMR videos generated by ECGFlowCMR. Five board-certified cardiologists, including one junior, three mid-level, and one senior expert, independently assessed 100 cine CMR videos in a binary real-versus-synthetic discrimination task. The evaluation set consisted of 50 ECGFlowCMR-generated videos and 50 real UKB CMR videos. All videos were randomly selected and shuffled before evaluation. The synthetic videos were generated from UKB ECGs covering diverse cardiac conditions, including 14 coronary artery disease (CAD), 11 cardiomyopathy (CM), 12 heart failure (HF), and 13 healthy cases. The real videos were sampled from the corresponding UKB CMR sequences to provide a balanced comparison across normal and pathological cases.

Table~\ref{tab:expert_comparison} reports expert-level discrimination accuracy with 95\% confidence intervals. ECGFlowCMR achieves an average accuracy of 0.514, close to the chance level of 0.5 and lower than the CardioNets baseline accuracy of 0.588 under the same evaluation protocol. Individual expert accuracies for ECGFlowCMR range from 0.40 to 0.60. Expert~3 obtains the highest accuracy of 0.60 with a 95\% confidence interval of [0.50, 0.69], while the other experts perform near chance level. Since lower real-versus-synthetic discrimination accuracy indicates that synthetic videos are more difficult to distinguish from real videos, these results suggest that ECGFlowCMR produces cine CMR sequences with stronger perceptual realism than the baseline generative model.

\end{document}